\documentclass[12pt]{article}

\usepackage[utf8]{inputenc}

\usepackage{amsmath}
\usepackage{amssymb}
\usepackage{amsthm}
\usepackage{color}
\usepackage{fullpage}
\usepackage{graphicx}
\usepackage{tikz}

\DeclareMathOperator*{\COarginf}{arg\,inf}
\DeclareMathOperator*{\COargmin}{arg\,min}

\newtheorem{proposition}{Proposition}
\newtheorem{example}{Example}
\newtheorem{remark}{Remark}
\newtheorem{definition}{Definition}

\usetikzlibrary{calc,arrows}

\newcommand{\COstate}{x}
\newcommand{\COstaterv}{X}
\newcommand{\COstatespace}{\mathcal{X}}
\newcommand{\COaction}{a}

\newcommand{\COactionspace}{\mathcal{A}}
\newcommand{\COdecision}{d}

\newcommand{\COdecisionspace}{\mathcal{D}}
\newcommand{\COloss}{\ell}
\newcommand{\COutility}{u}
\newcommand{\COinfo}{y}
\newcommand{\COinforv}{Y}
\newcommand{\COinfospace}{\mathcal{Y}}
\newcommand{\COexperiment}{e}

\newcommand{\COexperimentspace}{\mathcal{E}}
\newcommand{\COsuit}{s}
\newcommand{\COsone}{{\scriptsize \textcircled{1}}}
\newcommand{\COstwo}{{\scriptsize \textcircled{2}}}
\newcommand{\COsthree}{{\scriptsize \textcircled{3}}}
\newcommand{\COsoneintext}{\textcircled{1}}
\newcommand{\COstwointext}{\textcircled{2}}
\newcommand{\COsthreeintext}{\textcircled{3}}

\usepackage{mathtools}
\newcommand*{\COdefeq}{\coloneqq}
\newcommand*{\COqefed}{\eqqcolon}

\begin{document}

\title{Optimality Criteria for Probabilistic Numerical Methods}

\author{Chris.\ J.\ Oates$^{1,2}$, Jon Cockayne$^3$, Dennis Prangle$^1$, \\
T.~J.~Sullivan$^{4,5}$, Mark Girolami$^{6,2}$ \\
\small $^1$Newcastle University, UK \\
\small $^2$Alan Turing Institute, UK \\
\small $^3$University of Warwick, UK \\
\small $^4$Freie Universit\"at Berlin, DE \\
\small $^5$Zuse Institute Berlin, DE \\
\small $^6$University of Cambridge, UK}

\maketitle


\begin{abstract}
It is well understood that Bayesian decision theory and average case analysis are essentially identical.
However, if one is interested in performing \emph{uncertainty quantification} for a numerical task, it can be argued that \textcolor{black}{standard approaches from the decision-theoretic framework are neither appropriate nor sufficient. }
\textcolor{black}{Instead, we consider a particular optimality criterion from Bayesian experimental design and study its implied optimal information in the numerical context.}
This information is demonstrated to differ, in general, from the information that would be used in an average-case-optimal numerical method.
The explicit connection to Bayesian experimental design suggests several distinct regimes in which optimal probabilistic numerical methods can be developed.
\end{abstract}

\section{Introduction} \label{sec: introduction}

To fix notation, consider the task of approximating a quantity of interest $\phi \colon \COstatespace \rightarrow \Phi$ based on finite information provided through a map $\COinfo_\COexperiment \colon \COstatespace \rightarrow \COinfospace_\COexperiment$.
There may be several such maps with equivalent computational cost, and these are indexed by $\COexperiment \in \COexperimentspace$.
This abstract formulation covers most basic numerical tasks, including numerical integration, numerical optimisation and the numerical solution of a differential equation \cite{Wozniakowski2009}.
Thus a \emph{numerical method} is considered as a map $\COdecision_\COexperiment \colon \COinfospace_\COexperiment \rightarrow \Phi$ and the set of all possible numerical methods is denoted $\COdecisionspace_\COexperiment$.

The performance of a numerical method can be quantified in several ways, but a common scenario is for $\COstatespace$ and $\Phi$ to be normed spaces, in which case the \emph{average-case error} of a numerical method $\COdecision_\COexperiment$ can be defined as
\begin{align*}
\text{ACE}_p(\COexperiment , \COdecision_\COexperiment) & \COdefeq \left( \int \| \COdecision_\COexperiment(\COinfo_\COexperiment(\COstate)) - \phi(\COstate) \|_\Phi^p \, \mathrm{d} \pi_{\COstaterv}(\COstate) \right)^{1/p}
\end{align*}
where $p \in [1,\infty)$ and $\pi_{\COstaterv}$ is a Borel probability distribution on $\COstatespace$, to be specified.
Average-case analysis can be motivated by considering the state $\COstate$ to be the realisation of a random variable $\COstaterv \sim \pi_{\COstaterv}$ and then assessing the average performance of a numerical method \cite{Ritter2000}.

In \emph{information-based complexity} there is interest in optimal numerical methods, their associated optimal information and the tractability of the numerical task itself.
For instance, an \emph{average-case-optimal numerical method} is defined as
\begin{align}
\COdecision_\COexperiment^{\ast} & \in \textcolor{black}{\COargmin}_{\COdecision_\COexperiment \in \COdecisionspace_\COexperiment } \; \text{ACE}_p(\COexperiment , \COdecision_\COexperiment) \label{eq: aca optimal method}
\end{align}
and \emph{average-case-optimal information} is defined as
\begin{align}
\COexperiment^{\ast} & \in \textcolor{black}{\COargmin}_{\COexperiment \in \COexperimentspace } \; \text{ACE}_p(\COexperiment , \COdecision_\COexperiment^{\ast}) . \label{eq: aca optimal info}
\end{align}
See \cite{Ritter2000,Traub2003,Traub1991,Wozniakowski2009}.
\textcolor{black}{Here and in the sequel we simply assume all minima are realised, so \eqref{eq: aca optimal method} and \eqref{eq: aca optimal info} are presented as $\COargmin$ rather than $\COarginf$.}
In \cite{Kadane1985} and later \cite{Diaconis1988} it was observed that an average-case-optimal numerical method is mathematically identical to a \emph{Bayes rule} in Bayesian decision theory when $\pi_\COstaterv$ is interpreted as the \emph{prior} distribution (see Section~\ref{sec: BDT}).

There has been recent interest in the development of \emph{probabilistic numerical methods}, which treat the approximation of $\phi(\COstate)$ as a statistical estimation task \cite{Larkin1972,Hennig2015}:
\begin{definition}[\cite{Cockayne2017}; Defn. 2.2]
A \emph{probabilistic numerical method} is a map $\COdecision_\COexperiment \colon \COinfospace_\COexperiment \rightarrow \mathcal{P}_\Phi$, where $\mathcal{P}_\Phi$ is the set of all Borel probability distributions on $\Phi$.
\end{definition}
The motivation for probabilistic numerical methods is to provide formal uncertainty quantification for the quantity of interest.
Recall that the push-forward $\phi_\# \pi$ of a distribution $\pi$ is defined as $(\phi_\# \pi)(S) = \pi(\phi^{-1}(S))$ for all measurable $S$.
Let $\pi_{\COinforv | \COexperiment} \COdefeq (y_\COexperiment)_\# \pi_\COstaterv$ and denote the disintegration\footnote{If all random variables admit densities with respect to a reference measure then $\pi_{\COstaterv | \COinfo, \COexperiment}$ is the usual conditional distribution of $\COstaterv$ given $\COinfo_\COexperiment(\COstaterv)$. A formal treatment of disintegration of measure is presented in \cite{Chang1997}.} of $\pi_\COstaterv$ along the map $\COinfo_\COexperiment$ by $\{\pi_{\COstaterv | \COinfo, \COexperiment}\}_{\COinfo \in \COinfospace_\COexperiment}$.
\begin{definition}[\cite{Cockayne2017}; Defn. 2.5] \label{def: Bayesian}
A probabilistic numerical method $\COdecision_\COexperiment \colon \COinfospace_\COexperiment \rightarrow \mathcal{P}_\Phi$ is said to be \emph{Bayesian} with prior $\pi_\COstaterv$ if $\COdecision_\COexperiment(\COinfo) = \phi_\# \pi_{\COstaterv | \COinfo, \COexperiment}$ for $\pi_{\COinforv | \COexperiment}$-almost all $\COinfo \in \COinfospace_\COexperiment$.
\end{definition}
Thus the output of a Bayesian probabilistic numerical method is a \emph{posterior} distribution providing uncertainty quantification for the quantity of interest.
The foundations of Bayesian probabilistic numerical methods were established in \cite{Cockayne2017} and Bayesian methods (in the strict sense of Defn.~\ref{def: Bayesian}) have been studied for numerical integration \cite{Briol2018,Karvonen2018a}, global optimisation \cite{Hennig2013,Mahsereci2015,Mockus1989}, ordinary \cite{Wang2018} and partial \cite{Cockayne2017a,Owhadi2015a} differential equations and linear algebra \cite{Bartels2018,Cockayne2018,Hennig:2015hf}.
Moreover, these methods are starting to see practical application \cite{Chen2018,Oates2019,Pruher2018}.

The distributional output of a Bayesian probabilistic numerical method could of course be reduced to a point estimator for the quantity of interest, i.e.\ a (traditional) numerical method.
Standard calculations from Bayesian decision theory, for example following \cite{Kadane1985}, imply that the mean of $\phi_\# \pi_{\COstaterv | \COinfo, \COexperiment}$ is an average-case-optimal numerical method with $p = 2$ (see Section~\ref{subsec: Bayes acts}).
Conversely, \cite{Diaconis1988} illustrated that the trapezoidal rule for numerical integration is a Bayes rule, which is in turn an average-case-optimal numerical method, when $\pi_\COstaterv$ is the standard Weiner measure on $C(0,1)$.
Despite their elegance, these connections rather overlook the main objective of a probabilistic numerical method, which is to provide formal uncertainty quantification for the quantity of interest.
Indeed, to draw an analogy, it is well-understood that different strategies are required for the contrasting goals of parameter estimation and prediction in the statistical context \cite{OHagan1978}.
In particular, the notion of optimal information put forward in average-case analysis (equivalently, Bayesian decision theory) is not necessarily an appropriate criteria on which to base the design of a Bayesian probabilistic numerical method.

The aim of this chapter is therefore twofold:
First, we review connections between average-case-optimal numerical methods, average-case-optimal information and approaches to Bayesian experimental design from the statistical literature.
Second, we discuss how optimal information could be defined for a Bayesian probabilistic numerical method, where the goal is to perform uncertainty quantification as opposed to direct estimation of a quantity of interest.
In particular, we explore a particular criterion proposed in the recent work of \cite{Cockayne2017} that has the advantage of being straightforward to approximate.

\section{Bayesian Decision Theory} \label{sec: BDT}

In this section we present a succinct overview of statistical decision theory \cite{Berger1985}, recalling that this is mathematically identical to average-case analysis, as explained in \cite{Kadane1985}.
Throughout we use calligraphic font to represent (topological) state spaces, e.g.\ $\COstatespace$.
Upper-case letters are used to denote (Borel) random variables, e.g.\ $\COstaterv$.
For convenience, we identify $\COstaterv$ with a random variable on $\COstatespace$ whose distribution is denoted as $\pi_\COstaterv$.
The elements of a state space are denoted with lower-case letters, e.g.\ $\COstate \in \COstatespace$.

Consider again an index set $\COexperimentspace$.
In what follows we adopt the terminology of Bayesian decision theory and, in particular, refer to $\COexperiment \in \COexperimentspace$ as an \emph{experiment}.
Our presentation will be more general than in Section~\ref{sec: introduction} in three respects, which are all consequences of the broader context in which Bayesian decision-theoretic methods have been developed:

First, we associate to each experiment $\COexperiment \in \COexperimentspace$ and each state $\COstate \in \COstatespace$ a random variable $\COinforv_\COexperiment(\COstate)$ taking values in $\COinfospace_\COexperiment$.
The conditional distribution of this random variable is denoted $\pi_{\COinforv | \COstate,\COexperiment}$.
Often $\COinforv_\COexperiment(\COstate) = \COinfo_\COexperiment(\COstate)$ with probability one, where $\COinfo_\COexperiment(\COstate)$ is a deterministic function of the state $\COstate$, as in Section~\ref{sec: introduction}, in which case $\pi_{\COinforv | \COstate,\COexperiment} = \delta(\COinfo_\COexperiment(\COstate))$ is an atomic measure centred on $\COinfo_\COexperiment(\COstate) \in \COinfospace_\COexperiment$.
The more general formulation allows for the possibility that information about the state $\COstate$ is corrupted by noise.
Let $\pi_{\COinforv | \COexperiment}$ be the distribution defined through marginalisation as
\begin{equation*}
\pi_{\COinforv | \COexperiment}(S) = \iint 1_S(\COinfo) \, \mathrm{d}\pi_{\COinforv | \COstate, \COexperiment}(\COinfo) \, \mathrm{d}\pi_{\COstaterv}(\COstate) ,
\end{equation*}
where $1_S$ is the indicator function for the measurable set \textcolor{black}{$S \subset \COinfospace_\COexperiment$}.
The computational cost of each experiment is considered to be identical, but experiments differ in what information about $\COstate$ is provided in $\COinforv_\COexperiment(\COstate)$.

Second, we formulate the goal of Bayesian decision theory as the selection of an \emph{action} $\COaction$ from a specified set  $\COactionspace$.
Often, as in Section~\ref{sec: introduction}, the aim is to approximate $\phi(x)$ and the set of actions $\COactionspace$ is identical to $\Phi$.
The more general formulation allows for more compex policy and control strategies to be considered in the statistical context.
Correspondingly, a \emph{decision rule} $\COdecision_{\COexperiment} \in \COdecisionspace_\COexperiment$ is considered to be a function $\COdecision_{\COexperiment} \colon \COinfospace_{\COexperiment} \rightarrow \COactionspace$.

Third, we consider an arbitrary \emph{loss function} $\COloss \colon \COstatespace \times \COactionspace \rightarrow \mathbb{R}$.
This includes the case $\COloss(\COstate,\COaction) = \|\phi(\COstate) - \COaction\|_\Phi^p$ from Section~\ref{sec: introduction}.
The choice of the loss function should be informed by the reason why the experiment is being conducted and, in an estimation context, should reflect the quantities of interest.
See Chapter 2 of \cite{Berger1985} for further discussion of this point.

\begin{example}[Numerical integration] \label{ex: integration 1}
To illustrate the notation, consider the numerical task of integrating a continuous function $\COstate \in \COstatespace = C(0,1)$.
The aim here is to select an element $\COaction \in \COactionspace = \mathbb{R}$ which represents an approximation to the quantity of interest $\phi(\COstate) = \int_0^1 \COstate(t) \, \mathrm{d}t$.
Here the experiment set $\COexperimentspace$ consists of vectors $\COexperiment = [t_1,\dots,t_{n-1}]$, along with fixed endpoints $t_0 = 0$, $t_n = 1$, and $t_0 \leq \dots \leq t_n$, corresponding to maps $\COinfo_\COexperiment(\COstate) = [\COstate(t_0) , \dots , \COstate(t_n)]$.
One natural loss function is $\COloss(\COstate,\COaction) = (\phi(\COstate) - \COaction)^2$.
This problem has been well studied and will be used in an illustrative capacity in the sequel.
\end{example}

\subsection{Bayes Risk}

The framework of statistical decision theory allows comparison of different decision rules based on their \emph{Bayes risk}:
\begin{align}
\text{BR}(\COexperiment,\COdecision_\COexperiment) & \COdefeq \iint \COloss(\COstate , \COdecision_{\COexperiment}(\COinfo)) \, \mathrm{d}\pi_{\COinforv | \COstate,\COexperiment}(\COinfo) \, \mathrm{d}\pi_{\COstaterv}(\COstate) \label{eq: risk}
\end{align}
The Bayes risk quantifies the prior expected loss associated to a decision rule $\COdecision_\COexperiment$ and therefore forms a natural criterion for the selection of a decision rule in the Bayesian context.
We therefore define a \emph{Bayes rule} for an experiment $\COexperiment \in \COexperimentspace$ to be
\begin{align}
\COdecision_{\COexperiment}^{\ast} & \in \textcolor{black}{\COargmin}_{\COdecision_\COexperiment \in \COdecisionspace_\COexperiment} \text{BR}(\COexperiment,\COdecision_\COexperiment) . \label{eq: def Bayes rule}
\end{align}
In a restricted context\footnote{Indeed, as highlighted in \cite{Kadane1985}, Eq.~\eqref{eq: risk} is identical to $\text{ACE}_p(\COexperiment,\COdecision_\COexperiment)^p$ when $\pi_{\COinforv | \COstate,\COexperiment} = \delta(\COinfo_\COexperiment(\COstate))$, $\COactionspace = \Phi$ and $\COloss(\COstate,\COaction) = \| \phi(\COstate) - \COaction \|_\Phi^p$.} the definition of a Bayes rule is mathematically identical to the definition of an average-case-optimal numerical method in Eq.~\eqref{eq: aca optimal method}.
Similarly, we can define an optimal experiment
\begin{align*}
\COexperiment^{\ast} & \in \COexperimentspace_{\text{BDT}}^{\ast} \COdefeq \textcolor{black}{\COargmin}_{\COexperiment \in \COexperimentspace} \text{BR}(\COexperiment,\COdecision_\COexperiment^{\ast}) ,
\end{align*}
which is analogous to average case optimal information in Eq.~\eqref{eq: aca optimal info}.

\begin{example}[Numerical integration, continued] \label{ex: integration 2}
\cite{Suldin1959,Suldin1960} considered the standard Wiener measure $\pi_{\COstaterv}$, a Gaussian measure on $C(0,1)$ characterised by the property that for all $t, t' \in [0,1]$ we have $\int \COstate(t) \, \mathrm{d}\pi_\COstaterv(\COstate) = 0$ and $\int \COstate(t) \COstate(t') \, \mathrm{d}\pi_\COstaterv(\COstate) = \min(t,t')$.
In that work it was shown that for each experiment $\COexperiment = [t_1,\dots,t_{n-1}]$ there exists a unique Bayes rule
\begin{equation*}
\COdecision_\COexperiment^{\ast} = \frac{1}{2} \sum_{i=1}^n (\COstate(t_{i-1}) + \COstate(t_i)) (t_i - t_{i-1}) ,
\end{equation*}
which we recognise as the trapezoidal rule.
Moreover, it was shown that there is a unique optimal experiment with $t_i = \frac{i}{n}$.
Further contributions in this direction include \cite{Diaconis1988,Larkin1972,Larkin1974,Oettershagen2017,OHagan1991,OHagan1992}.
See \cite{Ritter2000} for a book-length treatment.
\end{example}

\begin{remark}[Admissibility]
It is important to note that other notions of optimality for decision rules, such as admissibility, need not coincide with the Bayesian notion of optimality.
A decision rule $\COdecision_\COexperiment \in \COdecisionspace_\COexperiment$ is called \emph{admissible} if there exists no $\COdecision_\COexperiment' \in \COdecisionspace_\COexperiment$ such that
\begin{align*}
\int \COloss(\COstate , \COdecision_\COexperiment'(\COinfo)) \, \mathrm{d}\pi_{\COinforv | \COstate,\COexperiment}(\COinfo) & \leq \int \COloss(\COstate , \COdecision_\COexperiment(\COinfo)) \, \mathrm{d}\pi_{\COinforv | \COstate,\COexperiment}(\COinfo)
\end{align*}
for all $\COstate \in \COstatespace$, with strict inequality for some $\COstate \in \COstatespace$.
The simplest illustration is estimation of $\COstate \in \mathbb{R}$ based on $\COinforv | \COstate \sim N(\COstate , 1)$ and with $\COloss(\COstate,\COstate') = (\COstate - \COstate')^2$, where an admissible decision rule is $\COdecision(\COinfo) = \COinfo$, but this is not a Bayes rule for any proper prior on $\mathbb{R}$.
\textcolor{black}{(Throughout this contribution, $N(\mu,\Sigma)$ denotes a Gaussian distribution with mean $\mu$ and covariance $\Sigma$.)}
Some results on when Bayes rules are, and are not, admissible can be found in the references provided in Section 8.4 of \cite{Berger1985}.
The famous result of \cite{Wald1947} demonstrated that, under certain conditions, all admissible decision rules are so-called \emph{generalised} Bayes rules, a definition in which improper priors are permitted.
For the purposes of uncertainty quantification, however, we wish to remain in the Bayesian framework and consider only Bayes rules that arise from, and can be understood in terms of, a probabilistic model.
\end{remark}

%
%

\subsection{Bayes Acts} \label{subsec: Bayes acts}

Following the definition of a Bayes rule it is reasonable to ask what actions a Bayes rule would select.
To this end, denote the set of \emph{Bayes acts} as
\begin{align*}
\COactionspace_\COexperiment^{\ast}(\COinfo_\COexperiment) & \COdefeq \textcolor{black}{\COargmin}_{\COaction \in \COactionspace} \int \COloss(\COstate , a) \, \mathrm{d} \pi_{\COstaterv | \COinfo, \COexperiment}(\COstate) .
\end{align*}
The proof of the following result is provided in Appendix~\ref{ap: norm vs ext proof}:
\begin{proposition} \label{prop: norm vs extensive}
A decision rule $\COdecision_\COexperiment^{\ast} \in \textcolor{black}{\COdecisionspace_\COexperiment}$ is a Bayes rule if and only if $\COdecision_{\COexperiment}^{\ast}(\COinfo) \in \COactionspace_\COexperiment^{\ast}(\COinfo)$ for $\pi_{\COinforv | \COexperiment}$-almost all $\COinfo \in \COinfospace_{\COexperiment}$.
\end{proposition}

Often the set $\COdecisionspace_\COexperiment$ will be infinite-dimensional, so that the search for a Bayes rule involves an optimisation problem which is infinite-dimensional.
However, the set $\COactionspace_\COexperiment$ is often finite-dimensional.
Hence, the action of a Bayes rule is something that can often be computed.

The next result is technical and will be needed later, to deduce that the mean of $\pi_{\COstaterv | \COinfo, \COexperiment}$ is a Bayes act for a certain family of loss functions.
Recall that a function $\phi \colon \COstatespace \rightarrow \mathbb{R}^m$ is called \emph{coercive} if for all $c > 0$ there exists a compact set $K \subset \mathcal{X}$ such that $\|\phi(\COstate)\|_2 \geq c$ for all $\COstate \in \mathcal{X} \setminus K$.
The proof of the following is provided in Appendix~\ref{ap: Bayes act proof}:
\begin{proposition} \label{prop: mean is Bayes act}
Consider $\COactionspace = \COstatespace = \mathbb{R}^d$.
Let $\COloss(\COstate,\COaction) = \| \phi(\COstate) - \phi(\COaction)\|_2^2$ where $\phi \colon \COstatespace \rightarrow \mathbb{R}^m$, $m \in \mathbb{N}$.
Assume that $\phi$ is twice continuously differentiable, that $\int \| \phi(\COstate) \|_{2} \, \mathrm{d} \pi_{\COstaterv | \COinfo,\COexperiment}(\COstate)$ is finite, and that the matrix
\begin{align}
\left[ \frac{\mathrm{d}\phi}{\mathrm{d}\COaction} \right]_{i,j} & = \frac{\partial}{\partial \COaction_j} \phi_i(\COaction) \label{eq: deriv matrix}
\end{align}
has full row rank at all $\COaction \in \COactionspace$.
Then any Bayes act $\COaction \in \COactionspace_\COexperiment^{\ast}(\COinfo_\COexperiment)$ satisfies
\begin{align}
\phi(\COaction) & = \int \phi(\COstate) \, \mathrm{d} \pi_{\COstaterv | \COinfo,\COexperiment}(\COstate). \label{eq: bayes act condition}
\end{align}
Moreover, if there exists a unique solution to Eq.~\eqref{eq: bayes act condition} and the function $\phi$ is coercive, then this solution is a Bayes act.
\end{proposition}

For simplicity we have presented this result in finite dimensions and for the standard Euclidean norm, but it can be naturally extended to infinite dimensions and to an arbitrary Hilbert space $\COstatespace = \COactionspace = \Phi$.

\begin{example}[Linear regression] \label{ex: regression}
Let $\pi_{\COstaterv} = N(\mu_0,\Sigma_0)$, $\COstaterv \in \mathbb{R}^d$, and $\pi_{\COinforv | \COstate,\COexperiment} = N(A_e \COstate , \Sigma)$, $\COinforv \in \mathbb{R}^n$, where the matrix $\Sigma_0$ is positive definite and the matrix $A_e \in \mathbb{R}^{n \times d}$ is determined by the choice of experiment $\COexperiment \in \COexperimentspace$.
Consider a loss $\COloss(\COstate,\COstate') = (\COstate - \COstate')^\top \Lambda (\COstate - \COstate')$ where $\Lambda$ is a positive semi-definite matrix with a square root $\Lambda^{\frac{1}{2}}$.
Then a Bayes decision rule $\COdecision_\COexperiment^{\ast}$ is defined through the Bayes act(s) $\COaction \in \mathbb{R}^d$ which, from Proposition~\ref{prop: mean is Bayes act}, satisfy $\Lambda^{\frac{1}{2}}\COaction = \Lambda^{\frac{1}{2}}\mu_{\COinfo,\COexperiment}$ where $\pi_{\COstaterv | \COinfo, \COexperiment} = N(\mu_{\COinfo,\COexperiment} , \Sigma_\COexperiment)$ and $\Sigma_\COexperiment = (A_e^\top A_e + \Sigma_0^{-1})^{-1}$, $\mu_{\COinfo,\COexperiment} = \Sigma_\COexperiment (A_e^\top \COinfo + \Sigma_0^{-1} \mu_0)$.
If $\Lambda$ is positive definite then it also follows from Proposition~\ref{prop: mean is Bayes act} that $\mu_{\COinfo , \COexperiment}$ is the unique Bayes act.
\end{example}

\begin{figure}[t!]
\centering
\begin{tikzpicture}[
        scale=1.5,
        axis/.style={very thick, ->, >=stealth'},
        important line/.style={thick},
        dashed line/.style={dashed, thick},
        every node/.style={color=black,}
     ]
    \coordinate (xint) at (2,0);
    \coordinate (end) at (3,0.8);
    \coordinate (end2) at (4,0);
    \coordinate (xintB) at (2.5,0);
    \coordinate (endB) at (3,1.5);
    \coordinate (end2B) at (3.5,0);
    \coordinate (act) at (3,0);
    \coordinate (truth) at (1,0);
    \coordinate (acthigh) at (3,1.5);
    \coordinate (truthhigh) at (1,1.5);
    \coordinate (loss) at (2,1.5);

    \begin{scope}
        \shade[top color=white, bottom color=blue, opacity=0.5]
            (xint) parabola [bend at end] (end) parabola [bend at start] (end2);
    \end{scope}
    \begin{scope}
        \shade[bottom color=white, top color=green, opacity=0.5]
            (xintB) parabola [bend at end] (endB) parabola [bend at start] (end2B);
    \end{scope}
    \draw[axis] (0,0)  -- (5.2,0) node(xline)[right] {$\COstatespace$};
    \draw[axis] (0,0) -- (0,1.7) node(yline)[above] {$\pi_{\COstaterv | \COinfo, \COexperiment}$};

    \draw[important line]
        (xint) parabola[bend at end] (end) parabola[bend at start] (end2)
        (xintB) parabola[bend at end] (endB) parabola[bend at start] (end2B);

    \fill[black] (act) circle (1pt) node[below] {$\COaction$};
    \fill[black] (truth) circle (1pt) node[below] {$\COstate$};

	\draw[dashed line] (truthhigh) -- (acthigh) node[pos=0.5, above]{$\COloss(\COstate,\COaction) = (\COstate - \COaction)^2$};
	\draw[dashed line] (truth) -- (truthhigh);
	\draw[dashed line] (act) -- (acthigh);
\end{tikzpicture}
\caption{Two posteriors ($\pi_{\COstaterv | \COinfo, \COexperiment}$; blue and green densities) can lead to the same Bayes act ($\COaction$; the posterior mean in the case of squared loss $\ell$) but provide very different uncertainty quantification for a quantity of interest ($\phi(\COstate)$).}
\label{fig: same mean same act}
\end{figure}
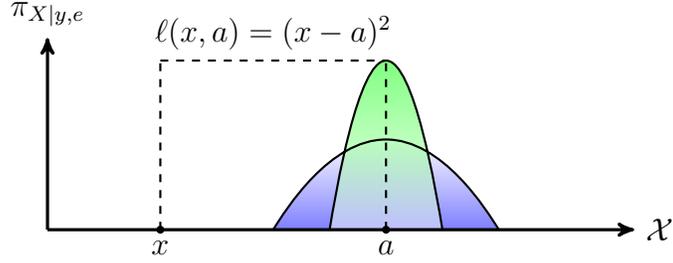

The explicit connection between Bayesian decision theory and average-case analysis \cite{Kadane1985} can be exploited to obtain information-based complexity results for Bayes rules and optimal experiments in the decision-theoretic context.
However, we recall that the differing goals of parameter estimation and uncertainty quantification need not lead to the same notions of an optimal experiment.
Indeed, two posteriors $\pi_{\COstaterv | \COinfo, \COexperiment}$ can lead to the same Bayes act but provide very different uncertainty quantification for a quantity of interest; see Figure~\ref{fig: same mean same act}.
To understand this point in detail, we turn to the more general framework of Bayesian experimental design in Section~\ref{sec: BED background}.

\section{Bayesian Experimental Design} \label{sec: BED background}

The term ``Bayesian experimental design'' has various historical usages, but in this work we are broadly consistent with the presentation of \cite{Chaloner1995}.
In this usage, Bayesian experimental design (BED) requires a \emph{utility} function $\COutility$ to be specified, such that $\COutility(\COexperiment, \COstate, \COinfo) \in \mathbb{R}$ represents the ``usefulness'' of the information $\COinfo \in \COinfospace_\COexperiment$ when the true state of nature is $\COstate \in \COstatespace$.
Armed with a utility function, we can construct a design criterion
\begin{align}
\text{BED}(\COexperiment) & \COdefeq - \iint \COutility(\COexperiment , \COstate , \COinfo) \, \mathrm{d}\pi_{\COinforv | \COstate, \COexperiment}(\COinfo) \, \mathrm{d}\pi_\COstaterv(\COstate) \label{eqn: BED def}
\end{align}
and the following notion of an optimal experiment is considered:
\begin{align*}
\COexperimentspace_{\text{BED}}^{\ast} & \COdefeq \textcolor{black}{\COargmin}_{\COexperiment \in \COexperimentspace} \text{BED}(\COexperiment) .
\end{align*}
See \cite{Clyde2001,Lindley1972,Mueller2005}.
In general optimal experimental designs will be analytically intractable, and in practice sophisticated numerical approaches to approximate the integral in Eq.~\eqref{eqn: BED def} and to perform the global multivariate optimisation may be needed, e.g. \cite{Amzal2006,Mueller2005,Overstall2018,Overstall2017}.
In an applied statistical context, the utility is something that itself may be elicited \cite{Wolfson1996} and robustness to perturbation of the utility can be investigated \cite{Ruggeri2005}.

The Bayesian experimental design framework is not concerned with selecting an action;
it concerns only selection of an experiment.
However, in terms of characterising an optimal experiment, the Bayesian experimental design framework is strictly more general than the Bayesian decision-theoretic approach, as we demonstrate next.

\subsection{Design Criteria via Decision Theory} \label{subsec: DCvDT}

First, we explain how the decision-theoretic approach can be recovered in the experimental design framework \cite{Lindley1972,Dawid1999}.
This is achieved by considering utilities $\COutility(\COexperiment, \COstate, \COinfo)$ which are independent of the state $\COstate$, in particular
\begin{align}
\COutility(\COexperiment , \COinfo) & = - \int \COloss(\COstate' ,  \COdecision_\COexperiment^{\ast}(\COinfo) ) \, \mathrm{d}\pi_{\COstaterv | \COinfo, \COexperiment}(\COstate') . \label{eq: BDT utility}
\end{align}
Under this choice, Eq.~\eqref{eqn: BED def} becomes
\begin{align*}
\text{BED}(\COexperiment) & = \iiint \COloss(\COstate', \COdecision_\COexperiment^{\ast}(\COinfo)) \, \mathrm{d}\pi_{\COstaterv | \COinfo, \COexperiment}(\COstate') \, \mathrm{d}\pi_{\COinforv | \COstate, \COexperiment}(\COinfo) \, \mathrm{d}\pi_{\COstaterv}(\COstate) \\
& = \iint \COloss(\COstate', \COdecision_\COexperiment^{\ast}(\COinfo)) \, \mathrm{d}\pi_{\COstaterv | \COinfo, \COexperiment}(\COstate') \, \mathrm{d}\pi_{\COinforv | \COexperiment}(\COinfo) \\
& = \iint \COloss(\COstate', \COdecision_\COexperiment^{\ast}(\COinfo)) \, \mathrm{d}\pi_{\COinforv | \COstate', \COexperiment}(\COinfo) \, \mathrm{d}\pi_{\COstaterv}(\COstate') \hspace{5pt} = \; \text{BR}(\COexperiment,\COdecision_\COexperiment^{\ast})
\end{align*}
and so $\COexperimentspace_{\text{BED}}^{\ast} = \COexperimentspace_{\text{BDT}}^{\ast}$.

To overcome the fact that $\text{BR}(\COexperiment,\COdecision_\COexperiment^{\ast})$ is analytically intractable in general, a suite of approximations specific to the decision-theoretic context have been developed.
Most approximations in the literature can be obtained by fixing a specific loss function $\COloss$ and then making a Gaussian approximation to the posterior;
see \cite{Dawid1999} for a more theoretical treatment.
In this section we briefly present the approach and, for concreteness, we suppose $\COstatespace = \mathbb{R}^m$ so that $\COstate$ is a column vector.
Therefore, we consider Gaussian approximations $\pi_{\COstaterv | \COinfo, \COexperiment} \approx N(\mu_{\COinfo,\COexperiment} , \Sigma_\COexperiment)$.
Note that the covariance $\Sigma_\COexperiment$ is considered to be independent of the information $\COinfo_\COexperiment$; a motivating example with this property was illustrated in Example~\ref{ex: regression}.
Then different choices of loss function $\COloss$ lead to different approximations, the co-called \emph{alphabet criteria}:
\begin{align*}
\text{BR}(\COexperiment,\COdecision_\COexperiment^{\ast}) & \approx \left\{ \begin{array}{ll} \text{tr}( \Lambda \Sigma_\COexperiment ) & \text{A-optimal} \\ \text{det}( \Lambda^{1/2} \Sigma_\COexperiment \Lambda^{1/2} ) & \text{D-optimal} \\ \vdots & \end{array} \right.
\end{align*}
for some positive semi-definite matrix $\Lambda$.
Armed with an explicit approximation to $\text{BR}(\COexperiment,\COdecision_\COexperiment^{\ast})$, global multivariate optimisation over $\COexperiment \in \COexperimentspace$ can be attempted; see the survey in \cite{Ryan2016}.

\begin{example}[Bayes A-, $c$- and E-optimality]
Consider a loss $\COloss(\COstate,\COstate') = \|\COstate - \COstate'\|_\Lambda^2$, where $\|\COstate\|_\Lambda \COdefeq \|\Lambda^{\frac{1}{2}} \COstate\|_2$ and $\Lambda$ is a positive semi-definite matrix with a square root $\Lambda^{\frac{1}{2}}$.
Suppose that the posterior $\pi_{\COstaterv | \COinfo,\COexperiment}$ is a Gaussian $N(\mu_{\COinfo,\COexperiment} , \Sigma_\COexperiment)$.
Then a Bayes decision rule $\COdecision_\COexperiment^{\ast}$ is defined through the Bayes act(s) $\COaction$, which satisfy $\Lambda^{\frac{1}{2}} \COaction = \Lambda^{\frac{1}{2}} \mu_{\COinfo,\COexperiment}$ due to Proposition~\ref{prop: mean is Bayes act}.
Now observe that $\int \COloss(\COstate , \COaction) \, \mathrm{d}\pi_{\COstaterv | \COinfo , \COexperiment}(\COstate) = \int (\COstate - \mu_{\COinfo,\COexperiment})^\top \Lambda (\COstate - \mu_{\COinfo,\COexperiment}) \, \mathrm{d}\pi_{\COstaterv | \COinfo , \COexperiment}(\COstate) = \text{\emph{tr}}(\Lambda \Sigma_\COexperiment)$, which is independent of $\COinfo_\COexperiment$.
It follows that $\text{\emph{BR}}(\COexperiment,\COdecision_\COexperiment^{\ast}) = \text{\emph{tr}}(\Lambda \Sigma_\COexperiment)$.
Selecting $\COexperiment$ to minimise $\text{\emph{tr}}(\Lambda \Sigma_\COexperiment)$, or $\text{\emph{tr}}(\Sigma_\COexperiment)$ in the common case where $\Lambda = I$, is called \emph{Bayes A-optimal} \cite{Owen1970,Brooks1972,Brooks1974,Brooks1976,Duncan1976,Brooks1977}.
In the special case where $\Lambda = c c^\top$ is a rank-1 matrix, the optimality criterion is called \emph{Bayes $c$-optimality} \cite{El-Krunz1991}.
Relatedly, minimisation of $\sup_{\|c\| = 1} \text{\emph{tr}}( c c^\top \Sigma_\COexperiment)$ is called \emph{Bayes E-optimality} \cite{Chaloner1984}.
\end{example}

\begin{example}[Bayes D-optimality]
Consider a loss $\COloss(\COstate,\COstate') = 0$ if $\|\COstate - \COstate'\|_\Lambda \leq \epsilon$ and, otherwise, $\COloss(\COstate,\COstate') = 1$.
Suppose that the posterior $\pi_{\COstaterv | \COinfo,\COexperiment}$ is a Gaussian $N(\mu_{\COinfo,\COexperiment} , \Sigma_\COexperiment)$.
A Bayes decision rule $\COdecision_\COexperiment^{\ast}$ is defined through the Bayes act(s), which one can verify include $\mu_{\COinfo,\COexperiment}$.
Now observe that $\int \COloss(\COstate , \mu_{\COinfo,\COexperiment}) \, \mathrm{d}\pi_{\COstaterv | \COinfo , \COexperiment}(\COstate) = \int 1_{\|\COstate - \mu_{\COinfo,\COexperiment}\|_\Lambda > \epsilon} \, \mathrm{d}\pi_{\COstaterv | \COinfo , \COexperiment}(\COstate)$, which is equal to one minus the probability that $\|Z\|_2 \leq \epsilon$ where $Z \sim N(0, \Lambda^{1/2} \Sigma_e \Lambda^{1/2})$.
For small $\epsilon$, this is $1 - O(c_d^{-1} \text{\emph{det}}(\Lambda^{1/2} \Sigma_\COexperiment \Lambda^{1/2})^{-d/2} \epsilon^d)$ where $c_d \COdefeq 2^{\frac{d}{2}} \Gamma(\frac{d}{2} + 1)$.
Note in particular that this is independent of $\COinfo_\COexperiment$.
It follows that $\text{\emph{BR}}(\COexperiment,\COdecision_\COexperiment^{\ast})$ is minimised when $\text{\emph{det}}(\Lambda^{1/2} \Sigma_\COexperiment \Lambda^{1/2})$ is minimised.
This criterion is called \emph{Bayes D-optimality} \cite{Tiao1976}.
\end{example}

The notions above can be naturally extended to infinite-dimensional state spaces; see e.g.\ \cite{Alexandrian2016}.
Given that the alphabet criteria are based on a Gaussian approximation of the posterior, it is incumbent on the analyst to attempt to verify the appropriateness of such approximations, for which a variety of diagnostics are available \cite{Clyde1993}.
The alphabet criteria can also be derived as prior expectations of classical experimental design criteria, usually based on the Fisher information, an approach that was called \emph{pseudo-Bayesian} in \cite{Ryan2016}.

\subsection{Design Criteria via Information Theory}

To perform Bayesian experimental design a decision-theoretic basis is not essential.
Indeed, other utilities --- often based on information theory --- have been proposed without reference to a decision-theoretic framework.
For instance, \cite{Lindley1956} proposed
\begin{align*}
\COutility(\COexperiment, \COinfo) & = D_{\text{KL}}( \pi_{\COstaterv | \COinfo, \COexperiment} \| \pi_{\COstaterv} ) ,
\end{align*}
where $D_{\text{KL}}$ denotes the Kullback-Leibler divergence (or \emph{information gain}) between the posterior and the prior.

This criterion is not explicitly motivated by estimation of a quantity of interest and may be better suited to tasks such as uncertainty quantification for the state $\COstate \in \COstatespace$.
However, it is interesting to note that the proposal of \cite{Lindley1956} can actually also be conceived from the decision-theoretic framework, albeit in a way that is non-standard \cite{Bernardo1979}.
Indeed, consider an action set $\COactionspace = \mathcal{P}_\COstatespace$, whose elements $\COaction \in \COactionspace$ are probability distributions on $\COstatespace$ and, for simplicity, existence of their Radon--Nikodym derivatives with respect to $\pi_\COstaterv$ is assumed.
Consider also a loss function of the form
\begin{align*}
\COloss(\COstate,\COaction) & = - \log\left( \frac{\mathrm{d}\COaction}{\mathrm{d}\pi_\COstaterv}(\COstate) \right) + f(\COstate)
\end{align*}
for arbitrary $f \colon \COstatespace \rightarrow \mathbb{R}$.
Such a loss function is known as a \emph{proper scoring rule}, since it can be shown that the unique Bayes act is to honestly report posterior belief; $\COaction = \pi_{\COstaterv | \COinfo, \COexperiment}$, see Theorem 2 of \cite{Bernardo1979}.
Eq.~\eqref{eq: BDT utility} implies, in the particular case of $f \equiv 0$, that
\begin{align*}
\COutility(\COexperiment , \COinfo) & = - \int \COloss(\COstate ,  \COdecision_\COexperiment^{\ast}(\COinfo) ) \, \mathrm{d}\pi_{\COstaterv | \COinfo, \COexperiment}(\COstate) \\
& = \int \log\left( \frac{\mathrm{d}\pi_{\COstaterv | \COinfo, \COexperiment}}{\mathrm{d}\pi_\COstaterv}(\COstate) \right) \, \mathrm{d}\pi_{\COstaterv | \COinfo, \COexperiment}(\COstate) \; \COqefed \; D_{\text{KL}}(\pi_{\COstaterv | \COinfo, \COexperiment} \| \pi_\COstaterv).
\end{align*}
The theory of proper scoring rules can be developed further and the reader is referred to \cite{Ehm2012}.
For other connections to information theory, we refer the reader to \cite{Hainy2014}.

Next we explicitly address the challenge of designing a utility function that is commensurate with the aims of a Bayesian probabilistic numerical method.

\section{Optimality Criteria for Probabilistic Numerical Methods}

The purpose of a Bayesian probabilistic numerical method is to provide formal uncertainty quantification for the state $\COstate \in \COstatespace$ or a derived quantity of interest $\phi(\COstate) \in \Phi$.
To build intuition, we immediately provide an example of such distributional output:

\begin{example}[Numerical integration, continued] \label{ex: integration 3}
Consider the numerical integration task, set out in Example~\ref{ex: integration 1}.
The output of a Bayesian probabilistic numerical method, with prior $\pi_\COstaterv$ equal to standard Weiner measure defined in Example~\ref{ex: integration 2} and information $\COinfo_\COexperiment(\COstate) = [\COstate(t_0),\dots,\COstate(t_n)]$, is
\begin{align*}
\COdecision_\COexperiment(\COinfo) = N \left( \sum_{i=1}^n \frac{(\COstate(t_{i-1}) + \COstate(t_i))}{2} (t_i - t_{i-1}) , \sum_{i=1}^n \frac{(t_i - t_{i-1})^3}{12} \right) .
\end{align*}
An overview of Bayesian probabilistic numerical methods for integration is provided in \cite{Briol2018}.
\end{example}
The distributional output can be naturally reduced to a point estimator; for instance the mean is recognised as the trapezoidal rule, a classical numerical method.
In this spirit of estimation, one can apply the alphabet criteria from Section~\ref{subsec: DCvDT}; the univariate Gaussian nature of the output implies that A-, $c$-, E- and D- optimal experiments $\COexperiment$ all seek to minimise the variance $\frac{1}{12} \sum_{i=1}^n (t_i - t_{i-1})^3$.
Thus a decision-theoretic perspective leads to an optimal experiment with uniformly spaced nodes $t_i = \frac{i}{n}$ .

However, the motivation for the development of probabilistic numerical methods is that the full distributional output provides formal uncertainty quantification for the quantity of interest.
This objective is related \emph{but not identical} to achieving strong performance at the estimation task.
Indeed, for such uncertainty quantification to be useful, we put forward three desiderata to guide the design of a probabilistic numerical method $\COdecision_\COexperiment$:
\begin{enumerate}
\item The output $\COdecision_\COexperiment$ should ``put mass close'' to the true quantity of interest $\phi$, in a sense to be specified.
\item The output $\COdecision_\COexperiment$ should be ``calibrated'', in the sense that the true value $\phi(\COstate)$ is well modelled as a random sample from the distributional output.
\item The search for an optimal experiment $\COexperiment \in \COexperimentspace$ should, practically, not be too difficult.
\end{enumerate}
Two important remarks are in order:
First, notice the tension between (1) and (2), since (1) prefers an atomic mass placed on a ``best'' estimate, whilst (2) encourages probability density to be spread out so that the support at least contains the true value of the quantity of interest.
Second, observe that for a \emph{Bayesian} probabilistic numerical method it is not possible in general to ensure (2) through suitable choice of experiment $\COexperiment \in \COexperimentspace$, since the output $\COdecision_\COexperiment$ depends crucially on the prior $\pi_\COstaterv$.
Hence, our focus in the remainder is on satisfying (1) and (3) for a Bayesian probabilistic numerical method, with (2) being assumed through suitable elicitation of $\pi_\COstaterv$.
A detailed discussion of prior elicitation was provided in the supplement of \cite{Cockayne2017} and sophisticated approaches to data-dependent elicitation have recently been developed, e.g. \cite{Jagadeeswaran2018}.

\begin{remark}[Dependence on $\pi_\COstaterv$]
\textcolor{black}{
The dependence of a Bayesian decision rule $\COdecision_\COexperiment$ on the prior $\pi_\COstaterv$ is a fundamental aspect of Bayesian decision theory, thus the difficulties just described with regard to (2) are also fundamental.
The alternative approach to decision theory pursued by Wald \cite{Wald1945} circumvents the elicitation of a prior $\pi_\COstaterv$ by casting the decision problem in an adversarial context, such that an adversary selects $\tilde{\pi}_\COstaterv$, which need not coincide with $\pi_\COstaterv$, from which $\COstaterv \sim \tilde{\pi}_\COstaterv$ is generated.
The decision-maker then adopts a Bayes rule $\COdecision_{\COexperiment}^*$, as in \eqref{eq: def Bayes rule}, based on a prior $\pi_\COstaterv$ that solves the minimax optimisation problem
$$
\min_{\pi_\COstaterv} \max_{\tilde{\pi}_\COstaterv} \iint \COloss(\COstate , \COdecision_{\COexperiment}^*(\COinfo)) \, \mathrm{d}\pi_{\COinforv | \COstate,\COexperiment}(\COinfo) \, \mathrm{d}\tilde{\pi}_{\COstaterv}(\COstate) .
$$
Wald's approach is considered in detail in the context of probabilistic numerical methods in \cite{Owhadi2015,Owhadi2015a}, and forms an interesting point of contrast to the present discussion.
}
\end{remark}

\subsection{The BPN Criterion} \label{subsec: BPN crit subsec}

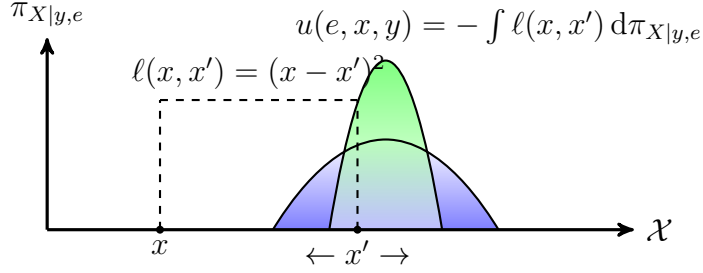
\begin{figure}[t!]
\centering
\begin{tikzpicture}[
        scale=1.5,
        axis/.style={very thick, ->, >=stealth'},
        important line/.style={thick},
        dashed line/.style={dashed, thick},
        every node/.style={color=black,}
     ]
    \coordinate (xint) at (2,0);
    \coordinate (end) at (3,0.8);
    \coordinate (end2) at (4,0);
    \coordinate (xintB) at (2.5,0);
    \coordinate (endB) at (3,1.5);
    \coordinate (end2B) at (3.5,0);
    \coordinate (act) at (2.75,0);
    \coordinate (truth) at (1,0);
    \coordinate (acthigh) at (2.75,1.15);
    \coordinate (truthhigh) at (1,1.15);
    \coordinate (loss) at (2,1.5);
    \coordinate (utility) at (4,1.8);

    \begin{scope}
        \shade[top color=white, bottom color=blue, opacity=0.5]
            (xint) parabola [bend at end] (end) parabola [bend at start] (end2);
    \end{scope}
    \begin{scope}
        \shade[bottom color=white, top color=green, opacity=0.5]
            (xintB) parabola [bend at end] (endB) parabola [bend at start] (end2B);
    \end{scope}
    \draw[axis] (0,0)  -- (5.2,0) node(xline)[right] {$\COstatespace$};
    \draw[axis] (0,0) -- (0,1.7) node(yline)[above] {$\pi_{\COstaterv | \COinfo, \COexperiment}$};

    \draw[important line]
        (xint) parabola[bend at end] (end) parabola[bend at start] (end2)
        (xintB) parabola[bend at end] (endB) parabola[bend at start] (end2B);

    \fill[black] (act) circle (1pt) node[below] {$\leftarrow \COstate' \rightarrow$};
    \fill[black] (truth) circle (1pt) node[below] {$\COstate$};

	\draw[dashed line] (truthhigh) -- (acthigh) node[pos=0.5, above]{$\COloss(\COstate,\COstate') = (\COstate - \COstate')^2$};
	\draw (utility) node {$\COutility(\COexperiment,\COstate,\COinfo) = - \int \COloss(\COstate,\COstate') \, \mathrm{d}\pi_{\COstaterv | \COinfo, \COexperiment}$};
	\draw[dashed line] (truth) -- (truthhigh);
	\draw[dashed line] (act) -- (acthigh);
\end{tikzpicture}
\caption{The $\text{BPN}$ criterion is motivated by the idea that a Bayesian probabilistic numerical method should concentrate its mass close to the true state ($\COstate$).
In that respect, the two posteriors ($\pi_{\COstaterv | \COinfo, \COexperiment}$; blue and green densities) lead to different utilities ($\COutility$) in general.
In this illustration, the more concentrated (green) posterior is preferred by the BPN criterion - in particular, the BPN criterion makes no attempt to ensure that the posterior is well calibrated.
(The calibration of the posterior depends chiefly on whether or not an appropriate prior $\pi_\COstaterv$ is elicited.)}
\label{fig: BPN illustration}
\end{figure}

In what follows we consider the action space $\COactionspace$ to be equal to the state space $\COstatespace$.
In that context, \cite{Cockayne2017} proposed the utility
\begin{align}
\COutility( \COexperiment , \COstate , \COinfo) & = - \int \COloss(\COstate, \COstate') \, \mathrm{d}\pi_{\COstaterv | \COinfo, \COexperiment}(\COstate') , \label{eq: our utility}
\end{align}
which differs from Eq.~\eqref{eq: BDT utility} in that it depends explicitly on the true state $\COstate$ and measures the posterior expected loss with respect to $\COstate$.
Thus two posteriors that give rise to the same Bayes act(s) need not give rise to the same utility function in Eq.~\eqref{eq: our utility}; see Figure~\ref{fig: BPN illustration}.
The associated experimental design criterion, denoted $\text{BPN}$, is a special case of Eq.~\eqref{eqn: BED def} with the above utility:
\begin{align}
\text{BPN}(\COexperiment) & := \iiint \COloss(\COstate , \COstate') \, \mathrm{d} \pi_{\COstaterv | \COinfo, \COexperiment}(\COstate') \, \mathrm{d} \pi_{\COinforv | \COstate, \COexperiment}(\COinfo) \, \mathrm{d} \pi_{\COstaterv}(\COstate). \label{eq: proposed}
\end{align}

\noindent To build intuition, for $\COstatespace$ a metric space with metric $d$ and $\COloss(\COstate,\COstate') = d(\COstate,\COstate')^p$, the inner integral
\begin{align}
\int \COloss(\COstate , \COstate') \, \mathrm{d} \pi_{\COstaterv | \COinfo, \COexperiment}(\COstate') & = D_{\text{W},p}(\delta(\COstate) , \pi_{\COstaterv | \COinfo , \COexperiment} ) \label{eq: Wass dist}
\end{align}
is the $p$\textsuperscript{th} Wasserstein distance between the posterior $\pi_{\COstaterv | \COinfo , \COexperiment}$ and an atom on the true state $\delta(\COstate)$.
Thus it is seen that desideratum (1) is satisfied.
Note also that that desideratum (3) is satisfied.
Indeed, in contrast to the decision-theoretic approach to experimental design, where Gaussian approximations such as the alphabet criteria are practically necessary for an optimal experiment to be computed (to circumvent the repeated calculation of a Bayes act), no Gaussian approximation is needed in order to numerically compute $\text{BPN}(\COexperiment)$.
Indeed, we can re-write the BPN criterion in Eq.~\eqref{eq: proposed} as
\begin{align}
\text{BPN}(\COexperiment) & = \int \underbrace{ \iint \COloss(\COstate , \COstate') \, \mathrm{d} \pi_{\COstaterv | \COinfo, \COexperiment}(\COstate') \, \mathrm{d} \pi_{\COstaterv | \COinfo, \COexperiment}(\COstate) }_{(\ast)} \mathrm{d} \pi_{\COinforv | \COexperiment}(\COinfo). \label{eq: BPN compute}
\end{align}
From this perspective it is clear that $\text{BPN}(\COexperiment)$ can be directly computed using a Monte Carlo method to sample from $\pi_{\COstaterv | \COinfo , \COexperiment}$ in order to approximate the inner integral $(\ast)$.
Although not closed form, we argue that the sophistication of modern Monte Carlo methods mean that desideratum (3) is essentially satisfied.
The intergal $(\ast)$ has been termed a \emph{distance-based information function} in \cite{Hainy2014} for the case $\COloss(\COstate,\COstate') = \|\phi(\COstate) - \phi(\COstate')\|_\Phi^2$.

\begin{example}[Numerical integration, continued]
Unpacking Example~\ref{ex: integration 3}, we note that $\pi_{\COstaterv | \COinfo, \COexperiment}$ is a collection of independent Brownian bridges $\COstaterv_i \colon [t_{i-1} , t_i] \rightarrow \mathbb{R}$ with $\COstaterv_i(t_{i-1}) = \COstate(t_{i-1})$, $\COstaterv_i(t_i) = \COstate(t_i)$ and covariance function $\text{\emph{cov}}(\COstaterv_i(t), \COstaterv_i(t')) = \frac{(t_i - t')(t-t_{i-1})}{(t_i - t_{i-1})}$, $t_{i-1} \leq t \leq t' \leq t_i$.
Through direct calculation we obtain the closed form $\text{\emph{BPN}}(\COexperiment) = \frac{1}{6} \sum_{i=1}^n (t_i - t_{i-1})^3$, which is easily minimised to obtain an optimal experiment by taking $t_i = \frac{i}{n}$.
That this should coincide with the BDT-optimal experiment is explained later in Proposition~\ref{thm: aca equivalence}.
\end{example}

In general the set of optimal experiments according to this criterion will be denoted
\begin{align*}
\COexperimentspace_{\text{BPN}}^{\ast} & \COdefeq \textcolor{black}{\COargmin}_{\COexperiment \in \COexperimentspace} \text{BPN}(\COexperiment) .
\end{align*}
For the remainder, our focus is on the properties of the BPN criterion and in particular we will ask, for this choice, whether $\COexperimentspace_{\text{BPN}}^{\ast} \stackrel{?}{=} \COexperimentspace_{\text{BDT}}^{\ast}$.

\subsection{Numerical Illustration} \label{sec: numerics}

It was explained in Section~\ref{subsec: BPN crit subsec} that the BPN criterion can be computed in circumstances that are quite general, as opposed to the BDT criteria for which Gaussian approximations are typically needed.
The aim of this section is to illustrate the use of the BPN criteria on a problem for which BDT cannot easily be applied.

Consider the canonical linear elliptic partial differential equation
\begin{align*}
- \Delta \COstate(t) & = f(t), &&  t \in D \\
\COstate(t) & = g(t), && t \in \partial D
\end{align*}
on $D \COdefeq (0,1)^2$.
Let $\COloss(\COstate,\COstate') = \left(\int (\COstate(t) - \COstate'(t))^p \, \mathrm{d}t\right)^{1/p}$ and let the prior $\pi_\COstaterv$ be Gaussian with $\int \COstate(t) \, \mathrm{d}\pi_\COstaterv(\COstate) = 0$, $\int \COstate(t) \COstate(t') \, \mathrm{d}\pi_\COstaterv(\COstate) = \exp(-\|t - t'\|_2^2)$.
An experiment $\COexperiment \in \COexperimentspace$ determines the set of locations $\{t_i\}_{i=1}^m \subset D$, $\{t_i\}_{i=m+1}^n \subset \partial D$ at which the functions $f$ and $g$, respectively, are evaluated.
To simplify the presentation we suppose that $m \ll n$ and focus on the information bottleneck, which is the placement of the points $\{t_i\}_{i=1}^m$ in $D$.
To minimise $\text{BPN}(\COexperiment)$ we recognised the triple integral in Eq.~\eqref{eq: BPN compute} as a single joint Gaussian integral and, to approximate this integral, we employed the standard Monte Carlo method.

\begin{figure}[t!]
\centering

\includegraphics[width = 0.25\textwidth]{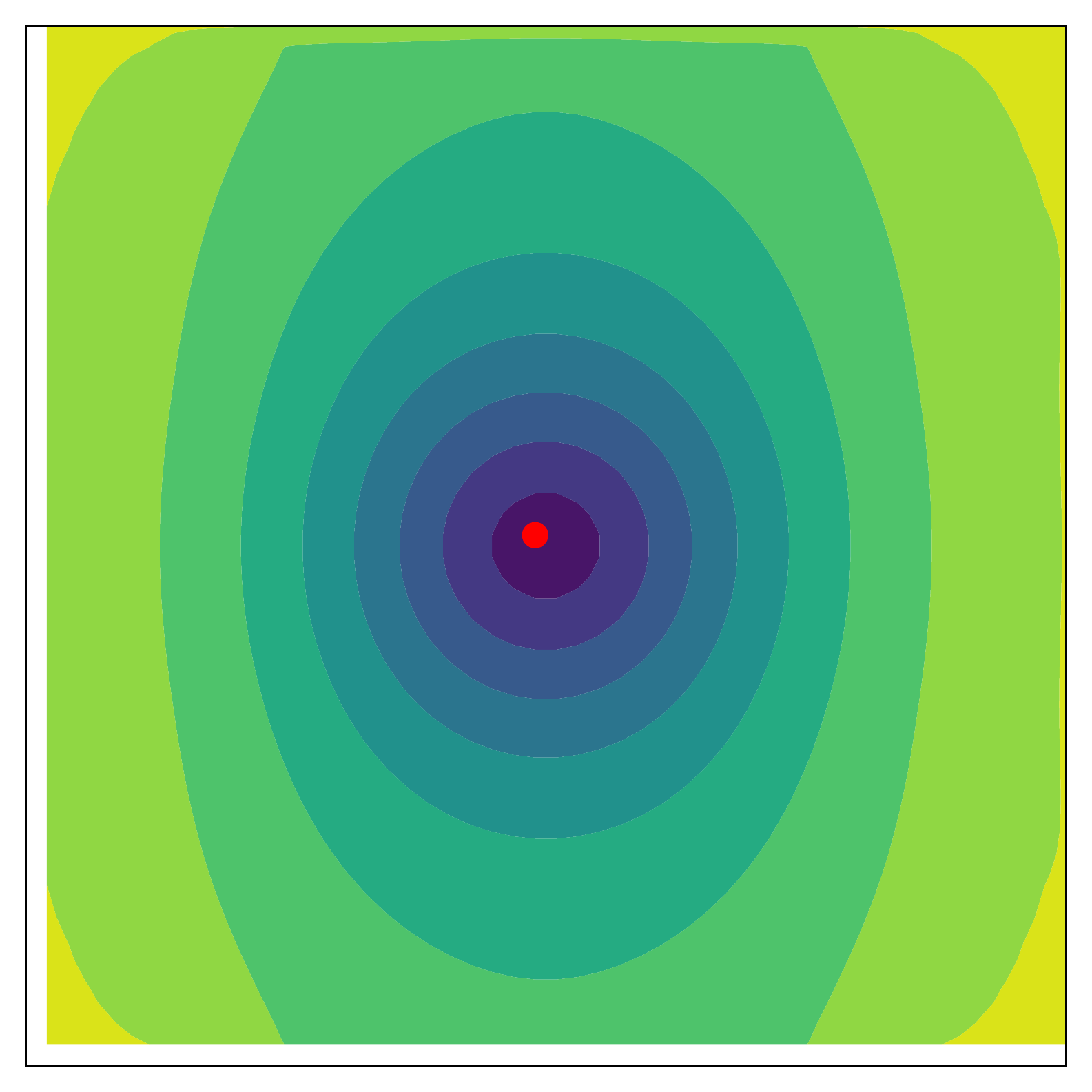}
\includegraphics[width = 0.25\textwidth]{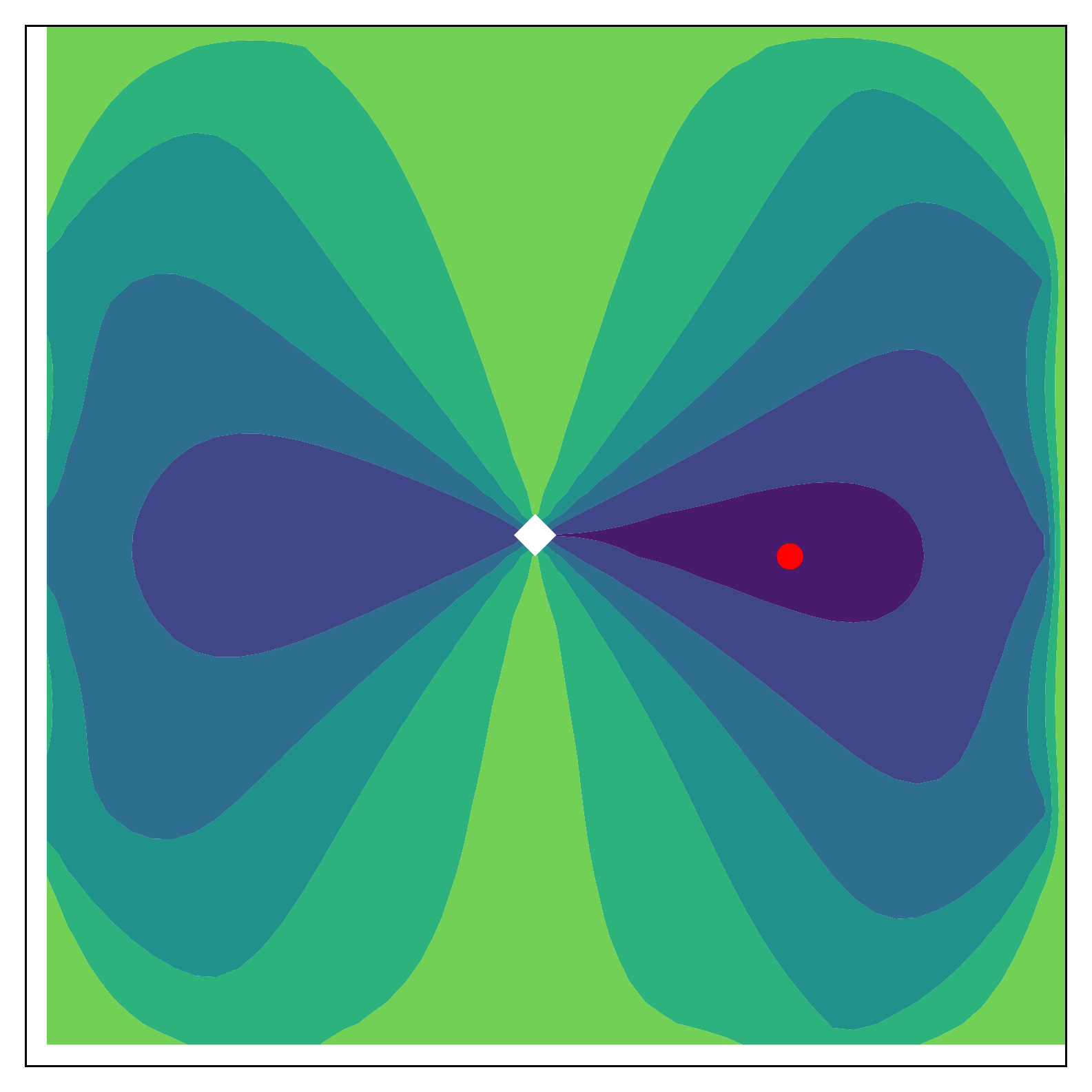}
\includegraphics[width = 0.25\textwidth]{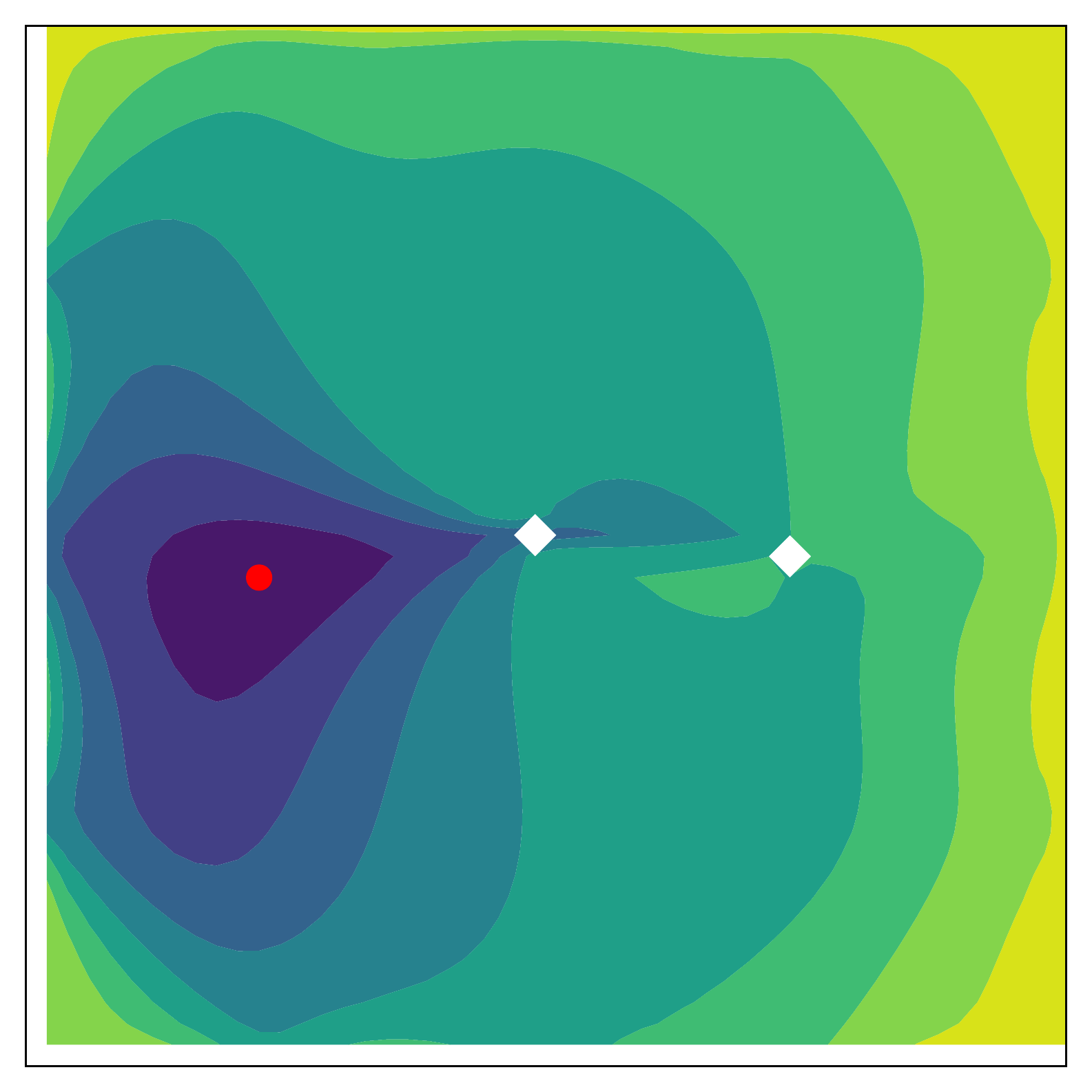}

\includegraphics[width = 0.25\textwidth]{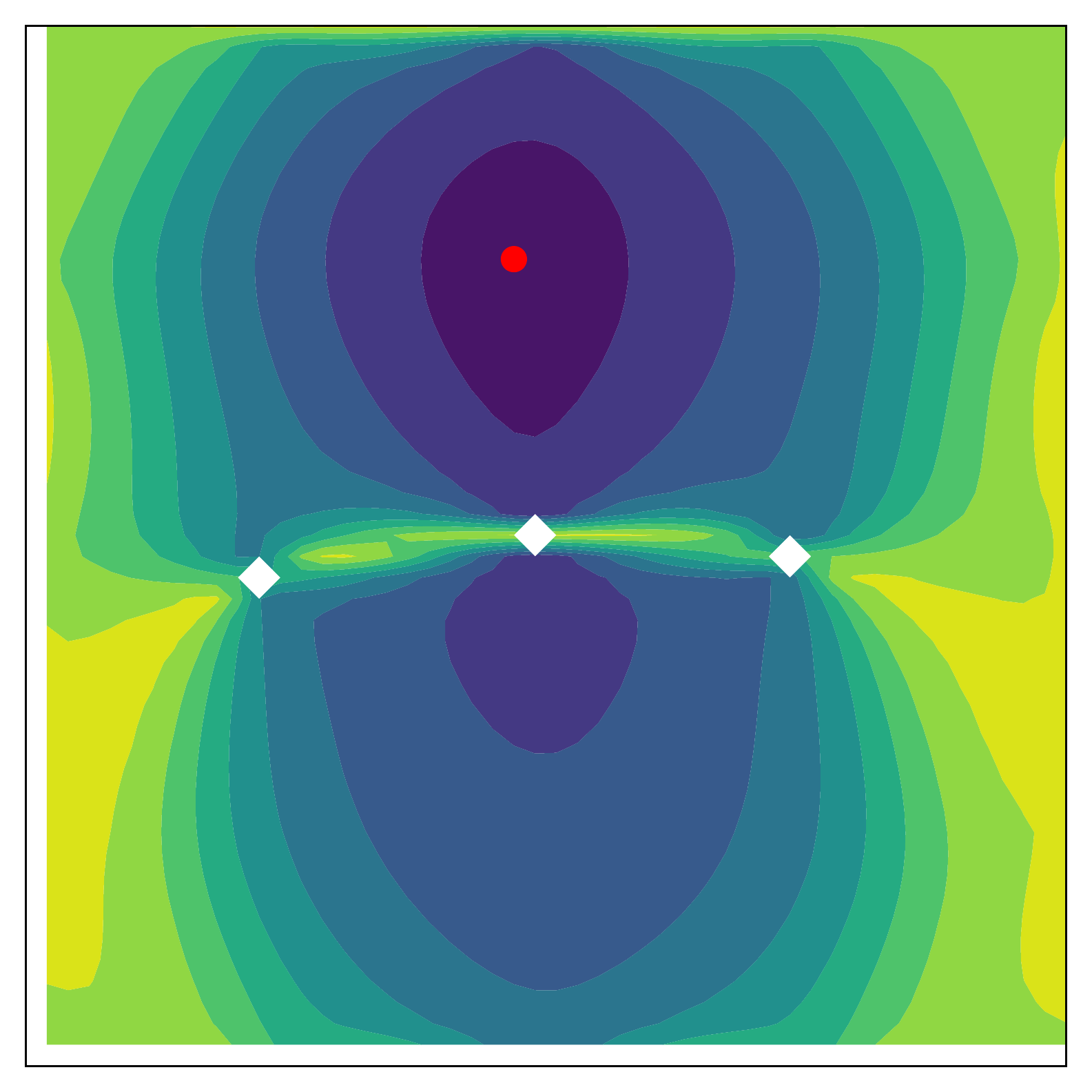}
\includegraphics[width = 0.25\textwidth]{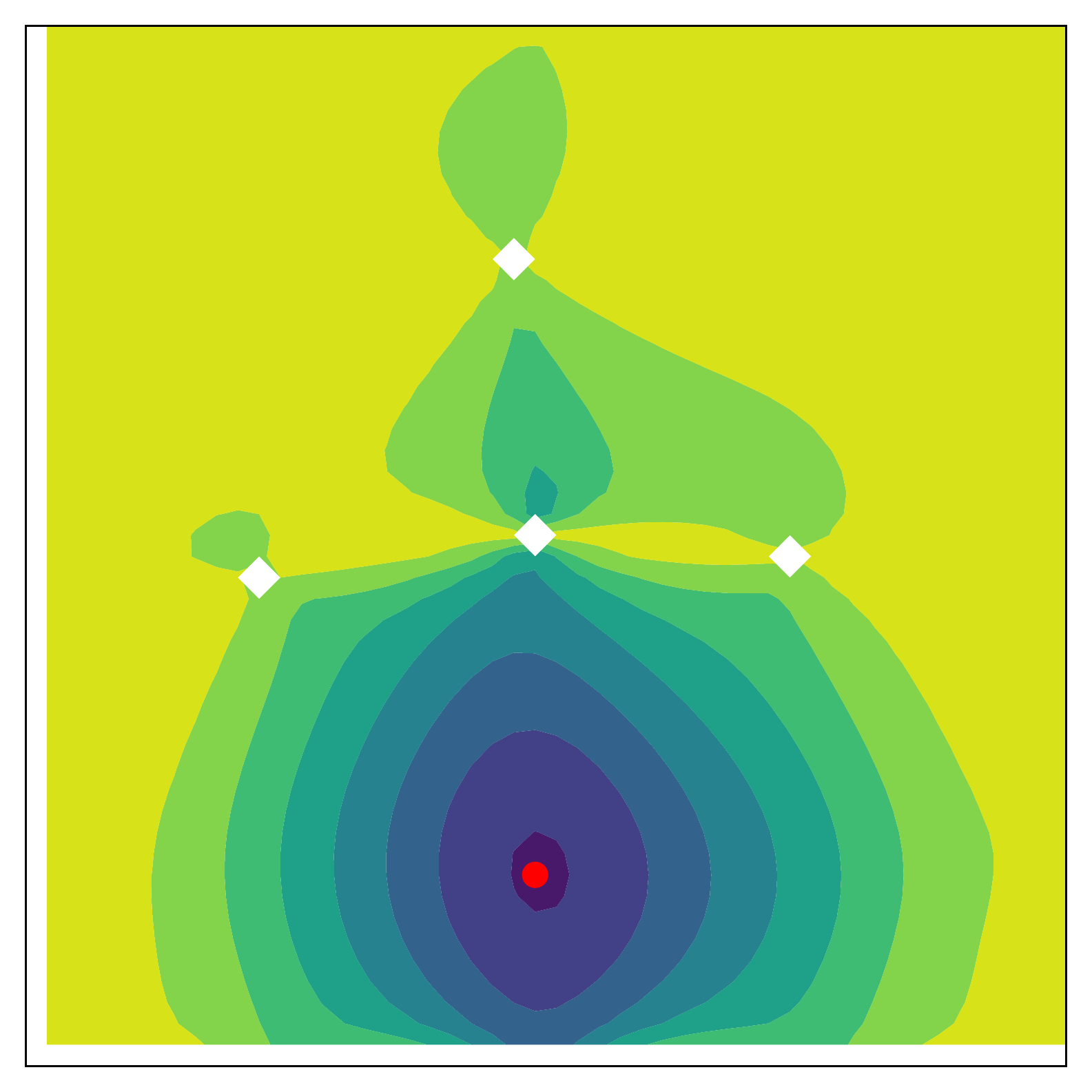}
\includegraphics[width = 0.25\textwidth]{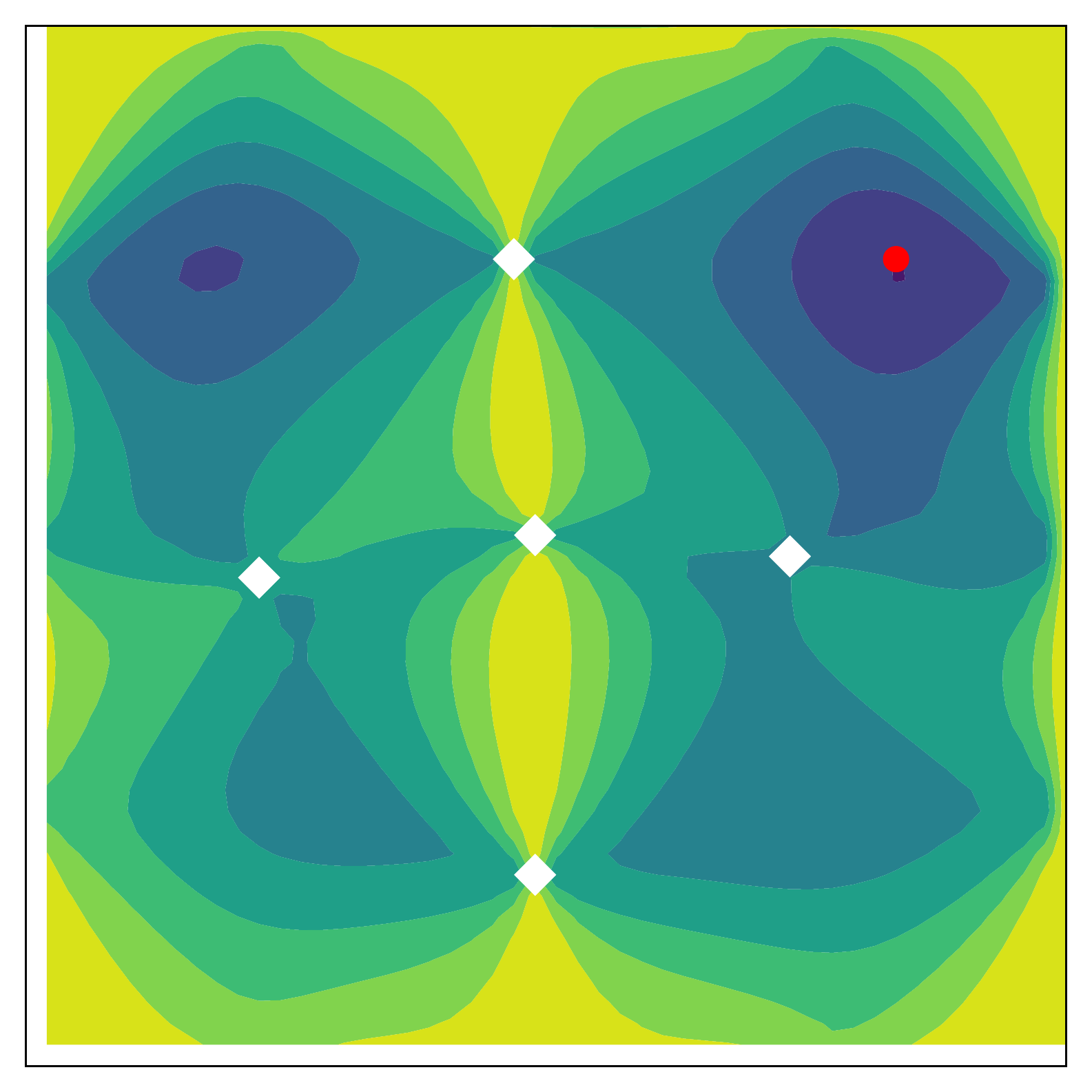}

\includegraphics[width = 0.25\textwidth]{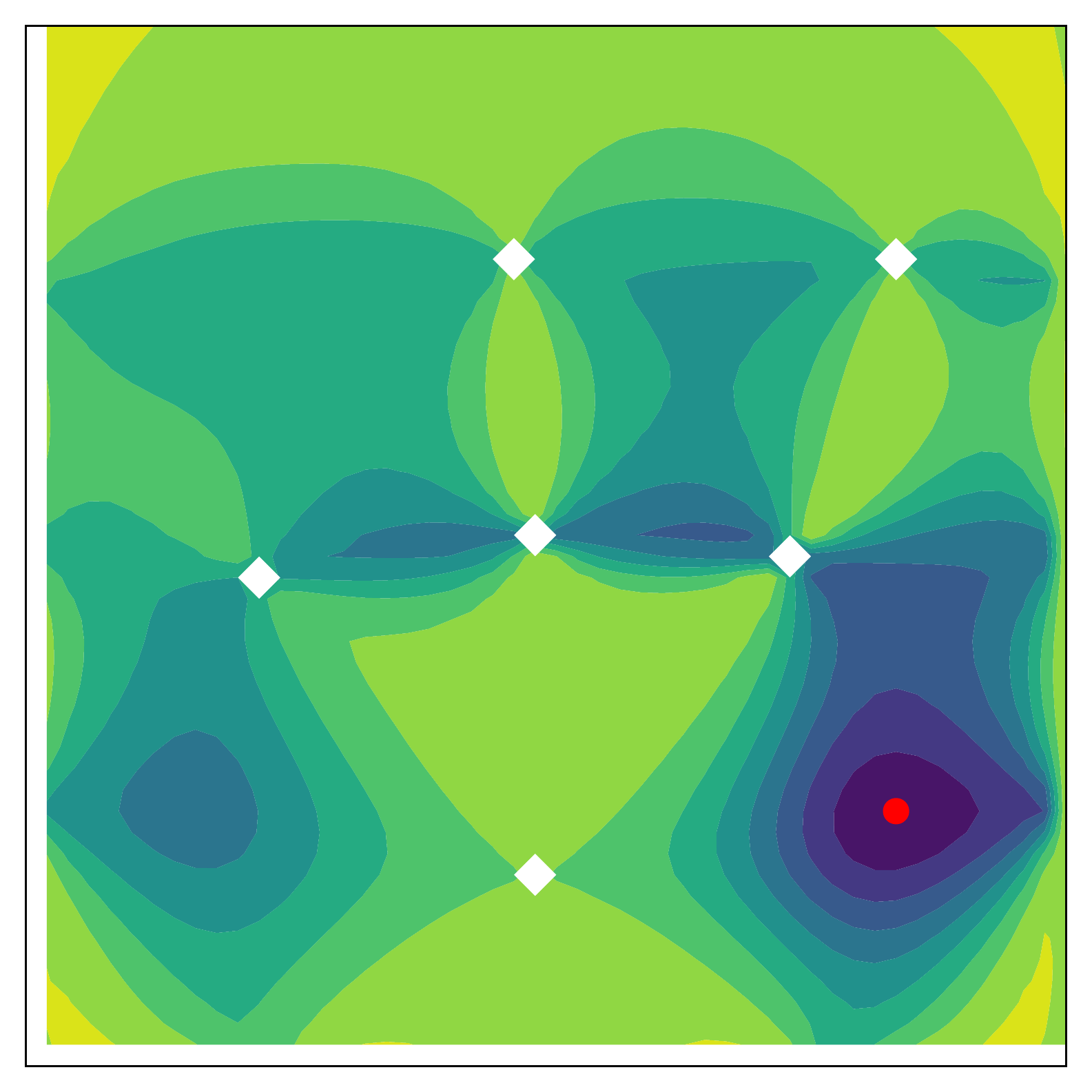}
\includegraphics[width = 0.25\textwidth]{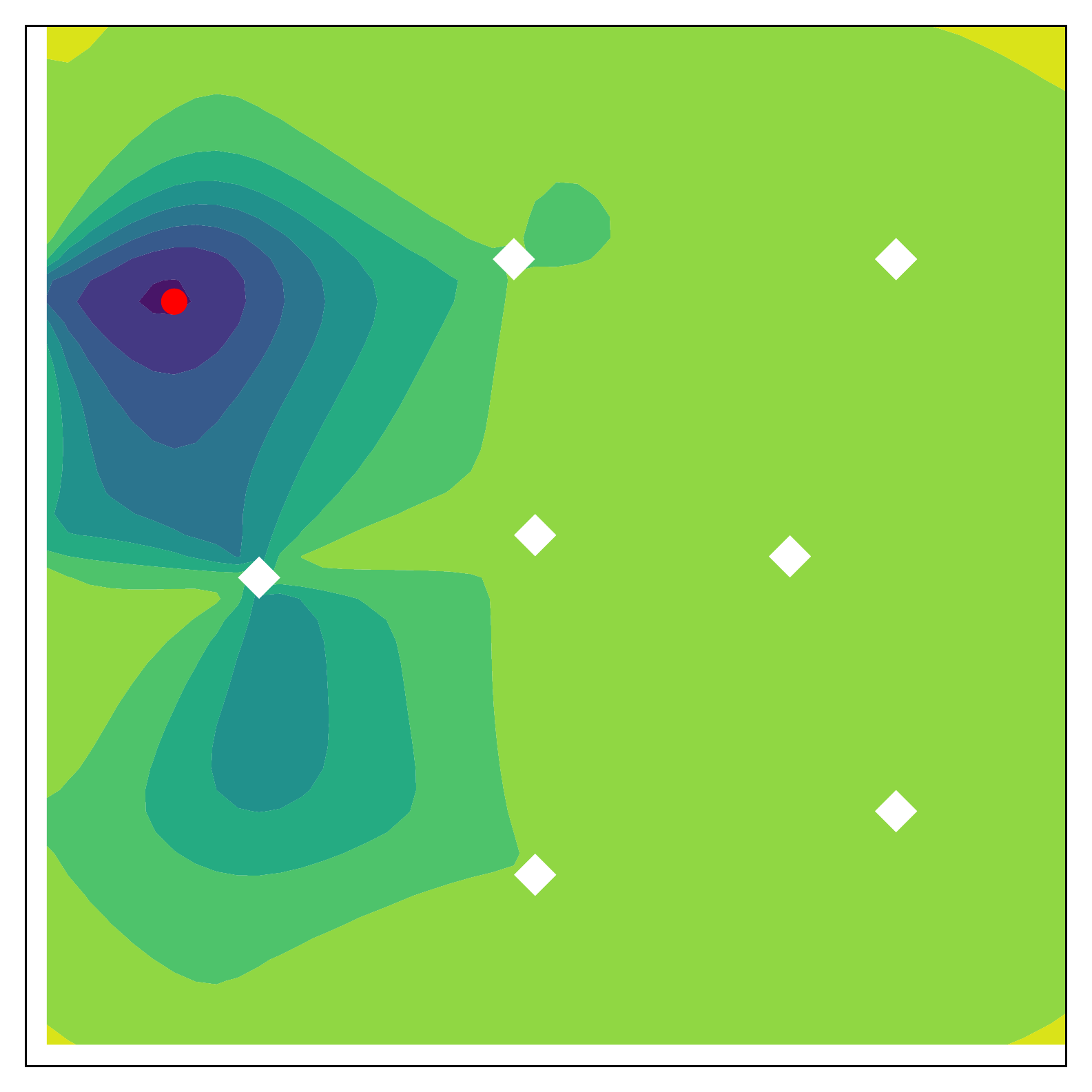}
\includegraphics[width = 0.25\textwidth]{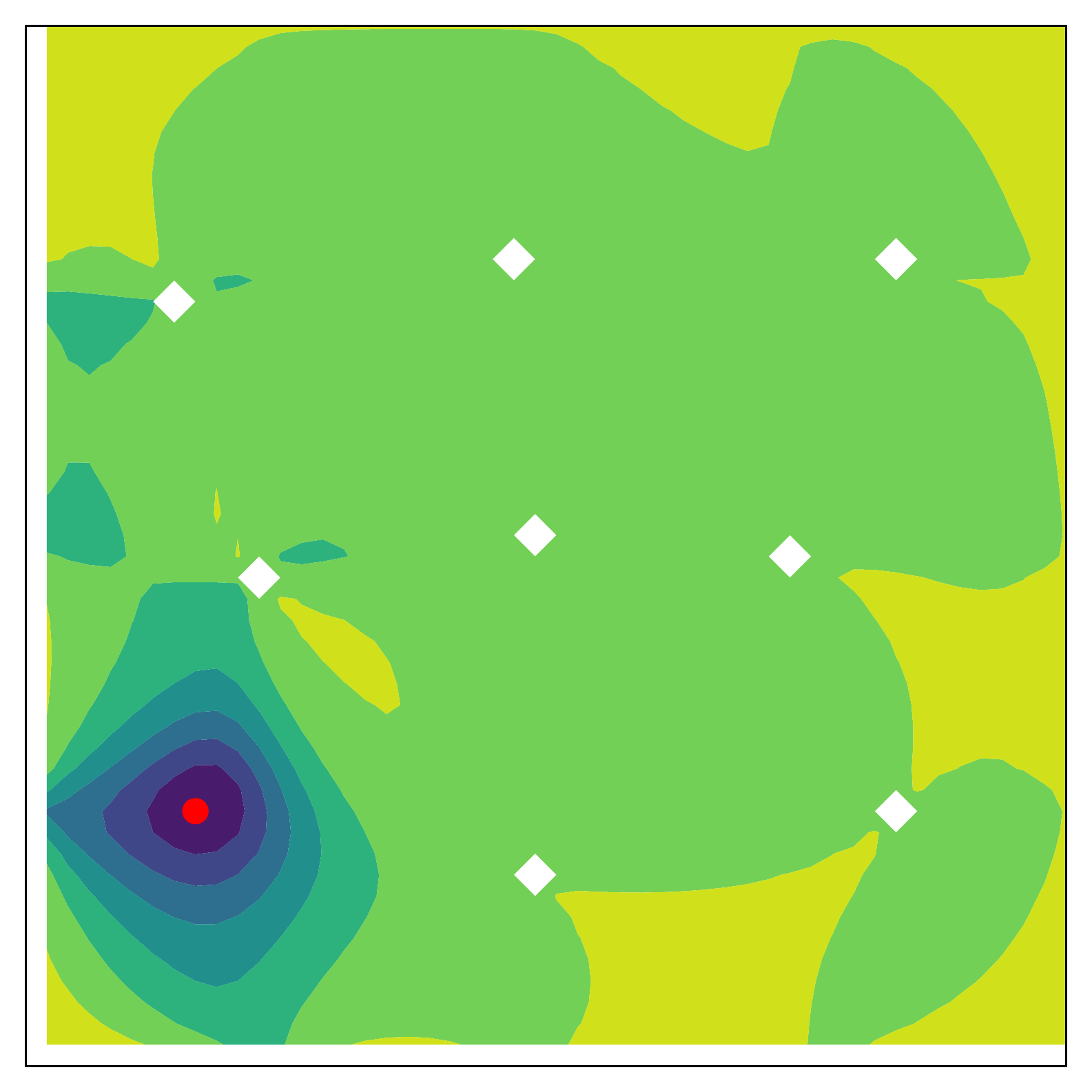}
\caption{Optimal experimental design with the BPN criterion, with $p = 2$ used.
(The filled contours depict $\text{BPN}(\COexperiment)$ as a function of $t_m$ where the lexicographic ordering of the panels corresponds to $m = 1,\dots,9$.
The white points represent the fixed values of $\{t_i\}_{i=1}^{m-1}$ and the red points represent the minimal value of $t_m$.)
}
\label{fig: p2}
\end{figure}

\begin{figure}[t!]
\centering

\includegraphics[width = 0.25\textwidth]{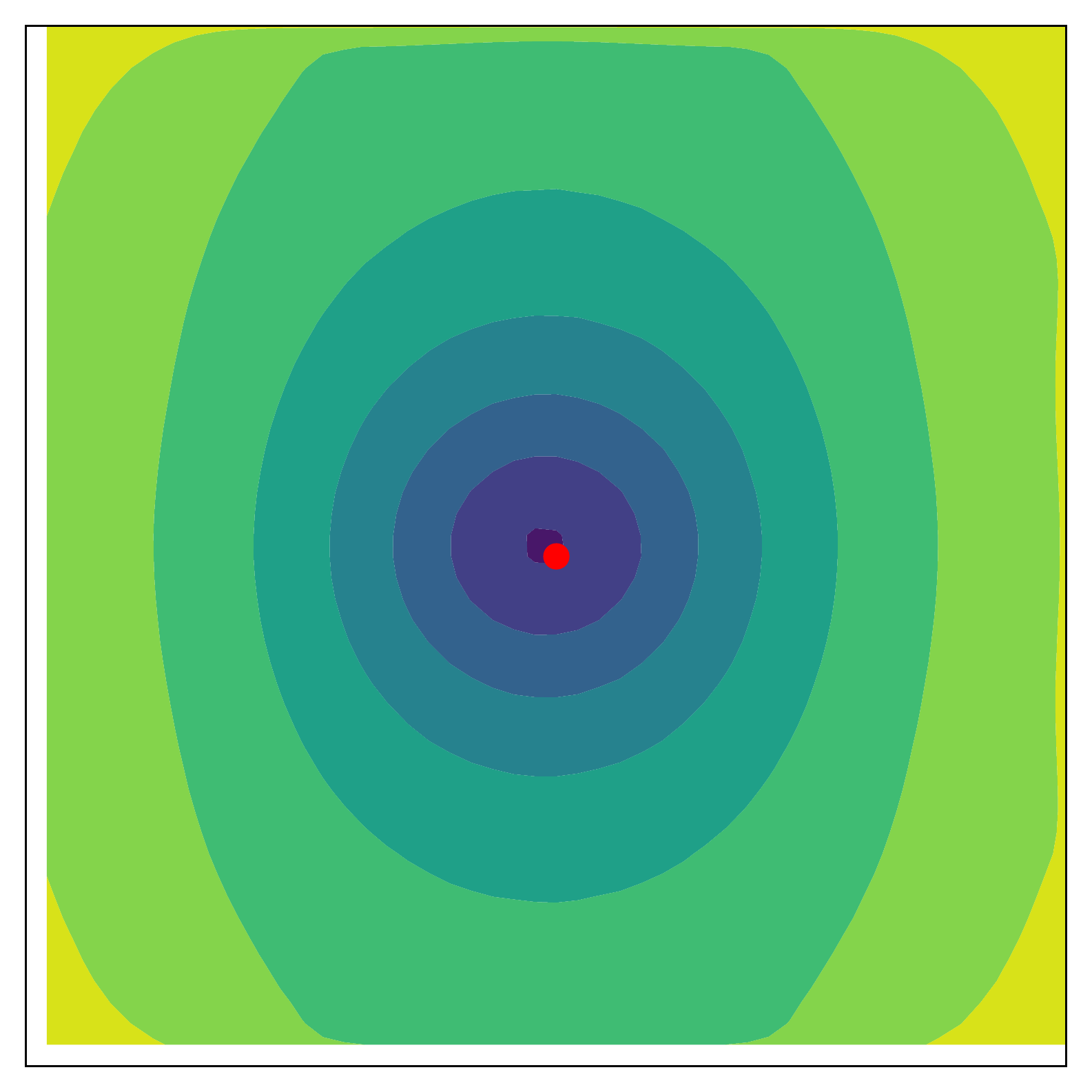}
\includegraphics[width = 0.25\textwidth]{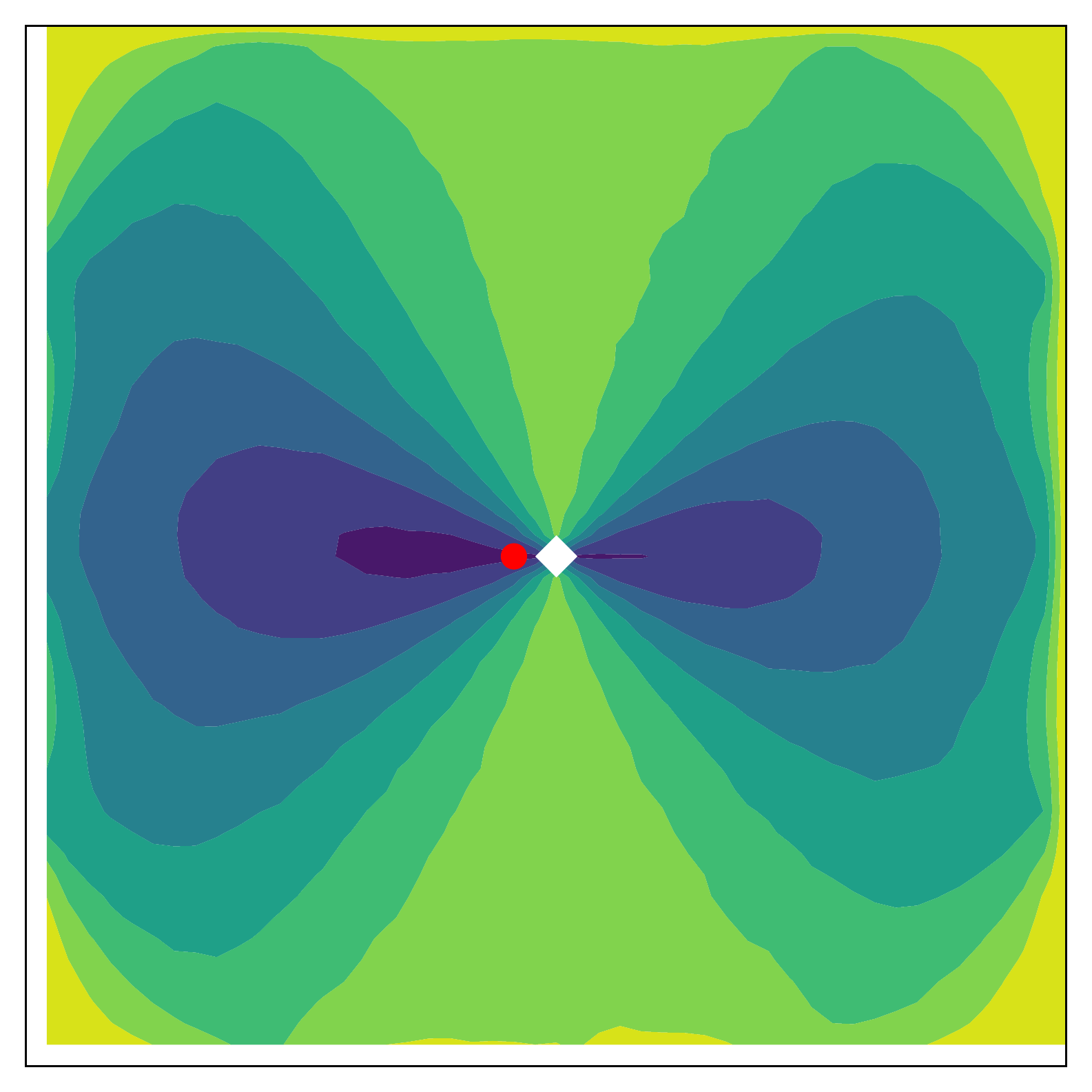}
\includegraphics[width = 0.25\textwidth]{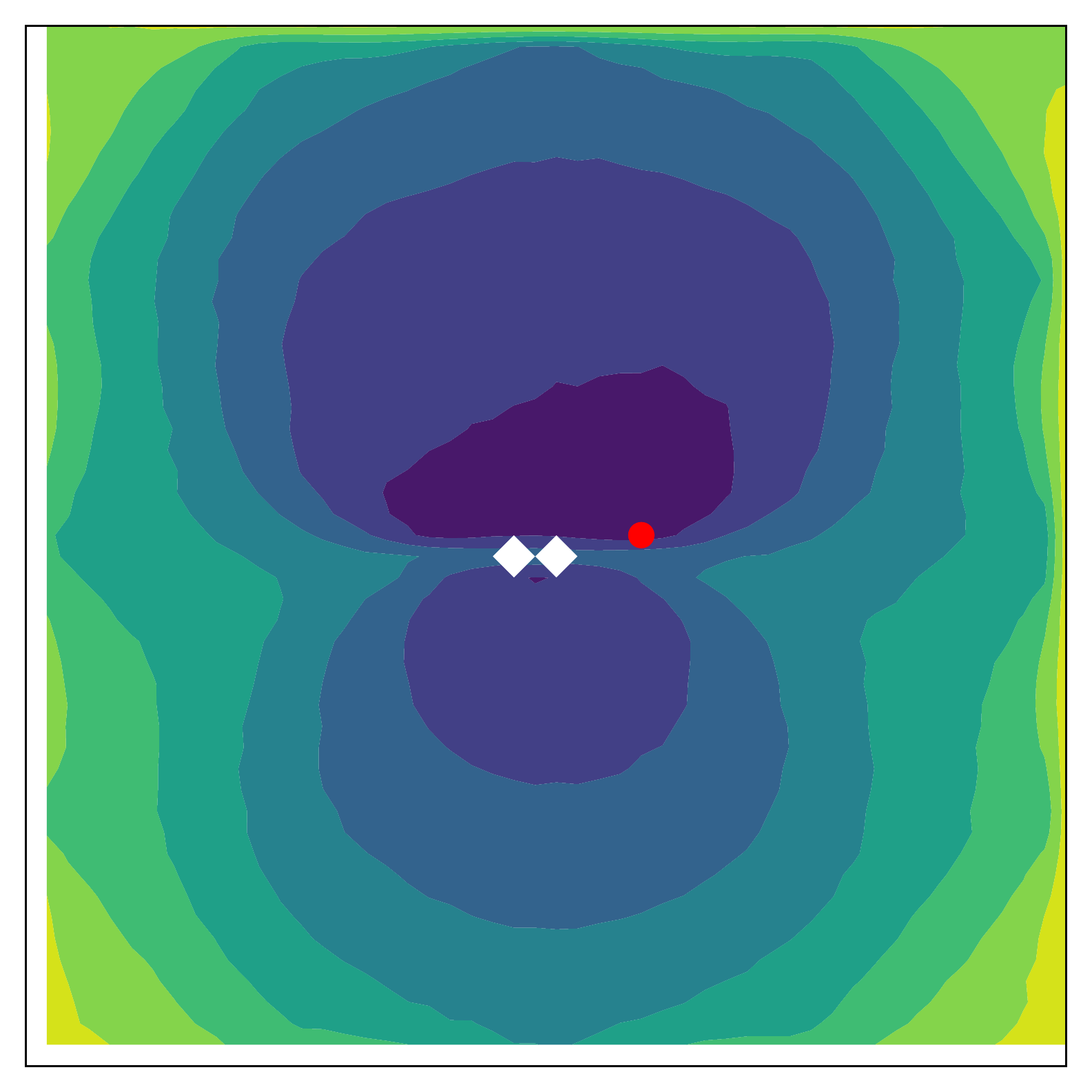}

\includegraphics[width = 0.25\textwidth]{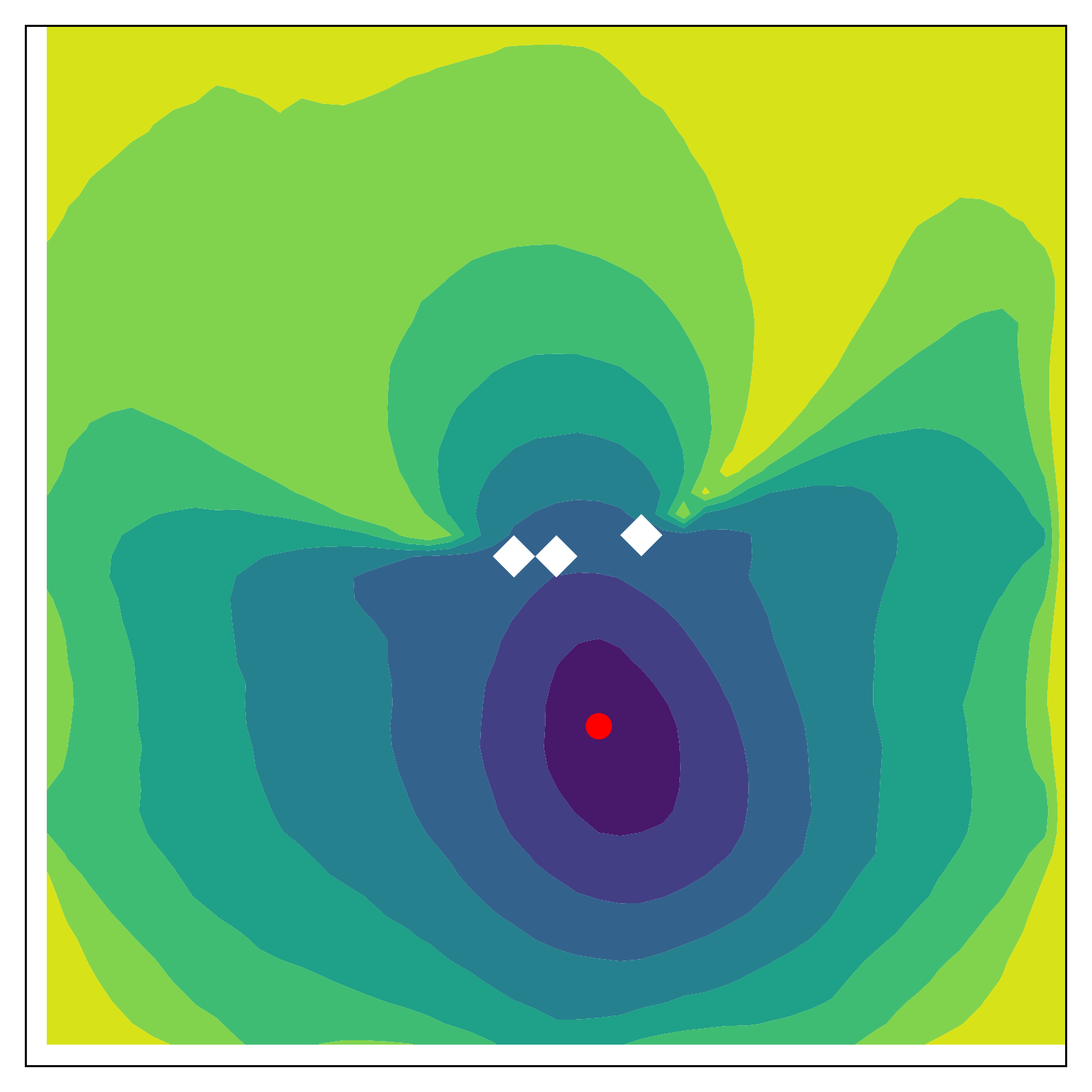}
\includegraphics[width = 0.25\textwidth]{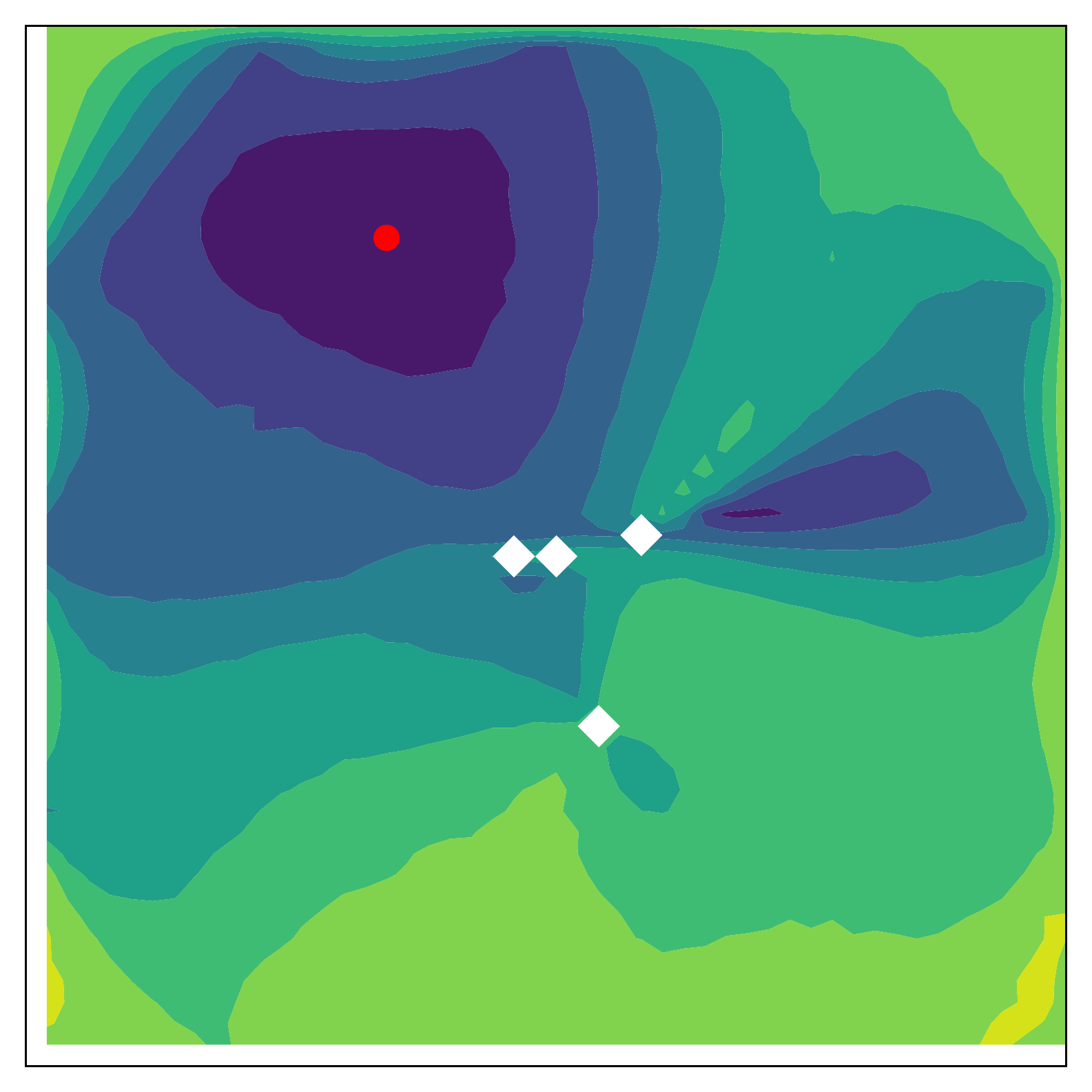}
\includegraphics[width = 0.25\textwidth]{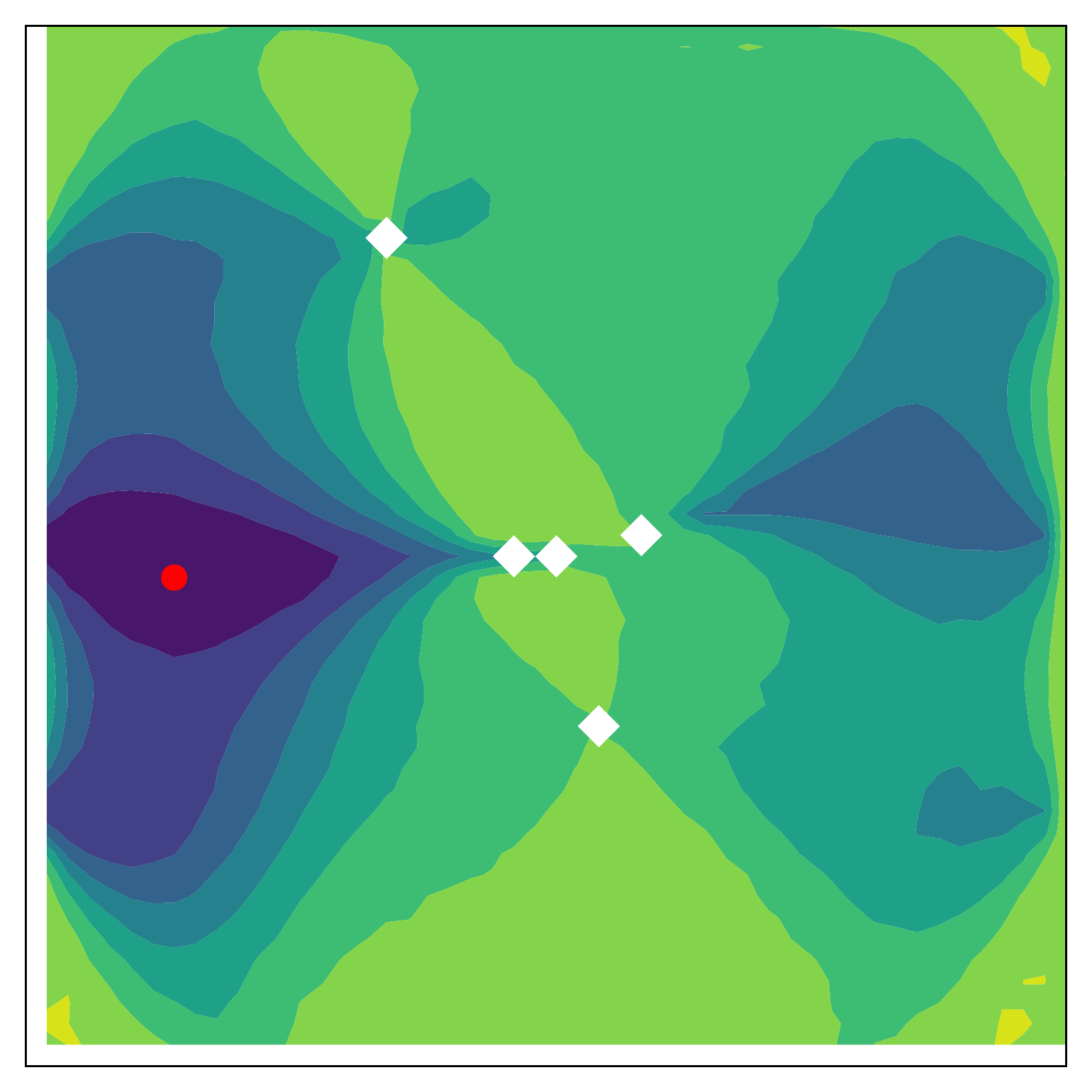}

\includegraphics[width = 0.25\textwidth]{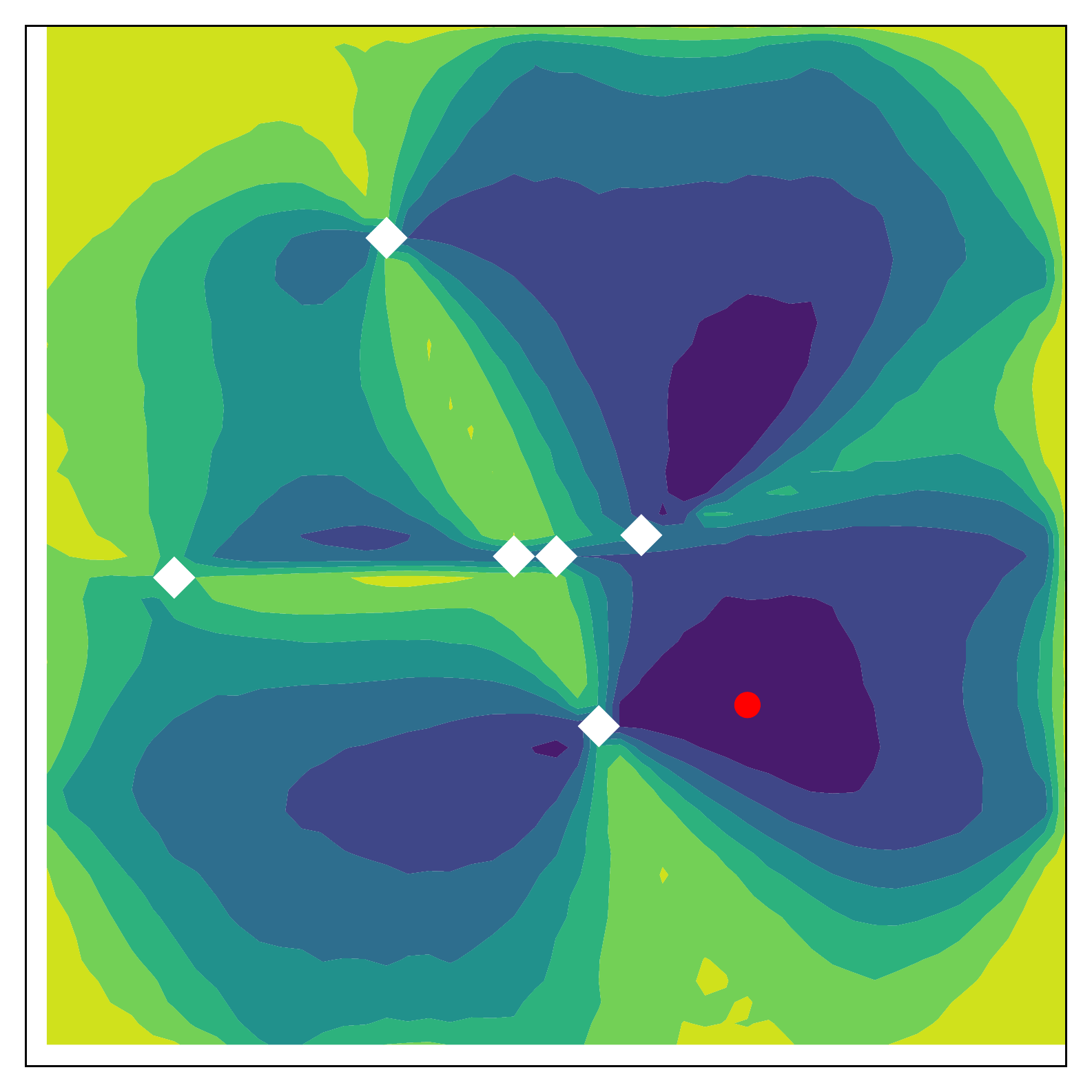}
\includegraphics[width = 0.25\textwidth]{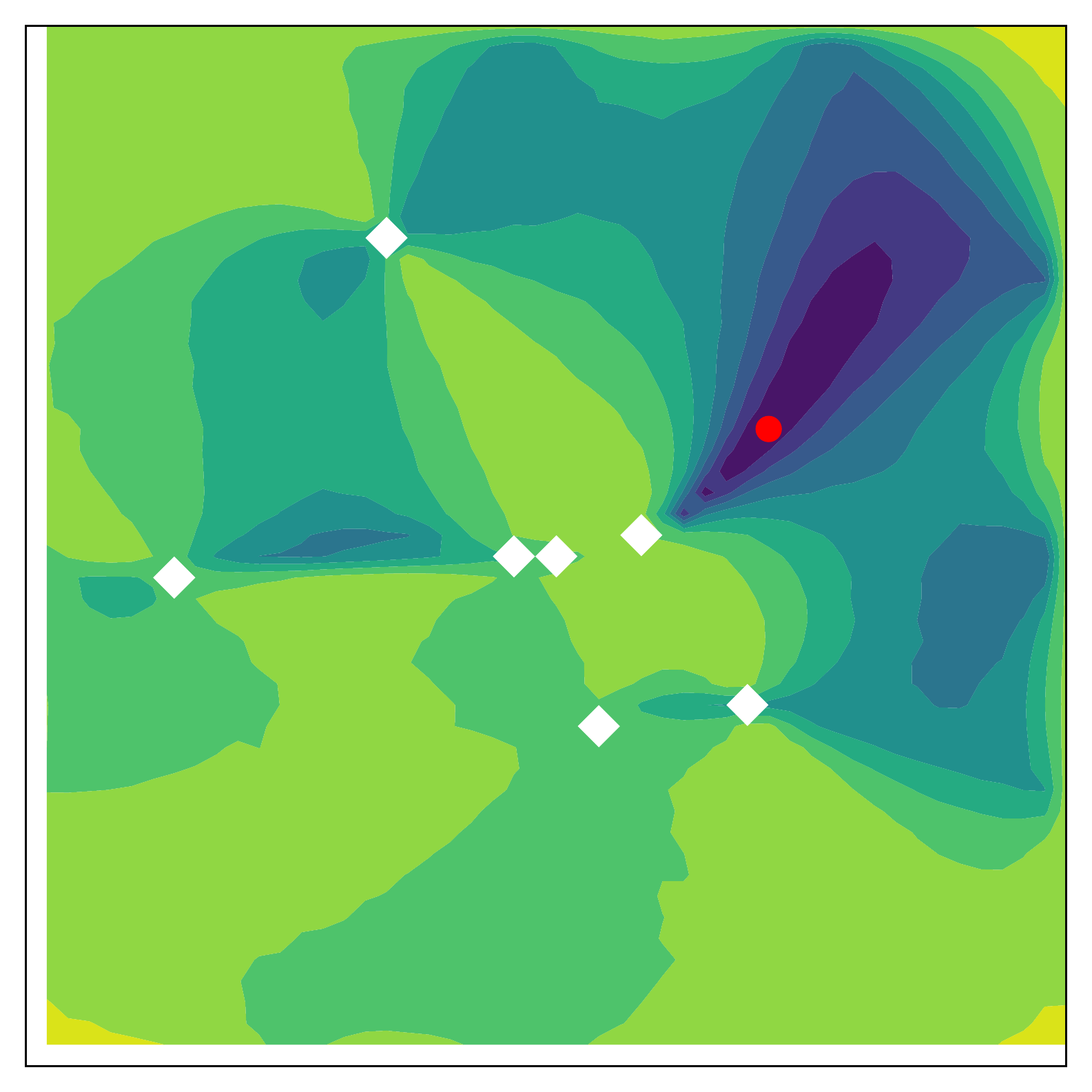}
\includegraphics[width = 0.25\textwidth]{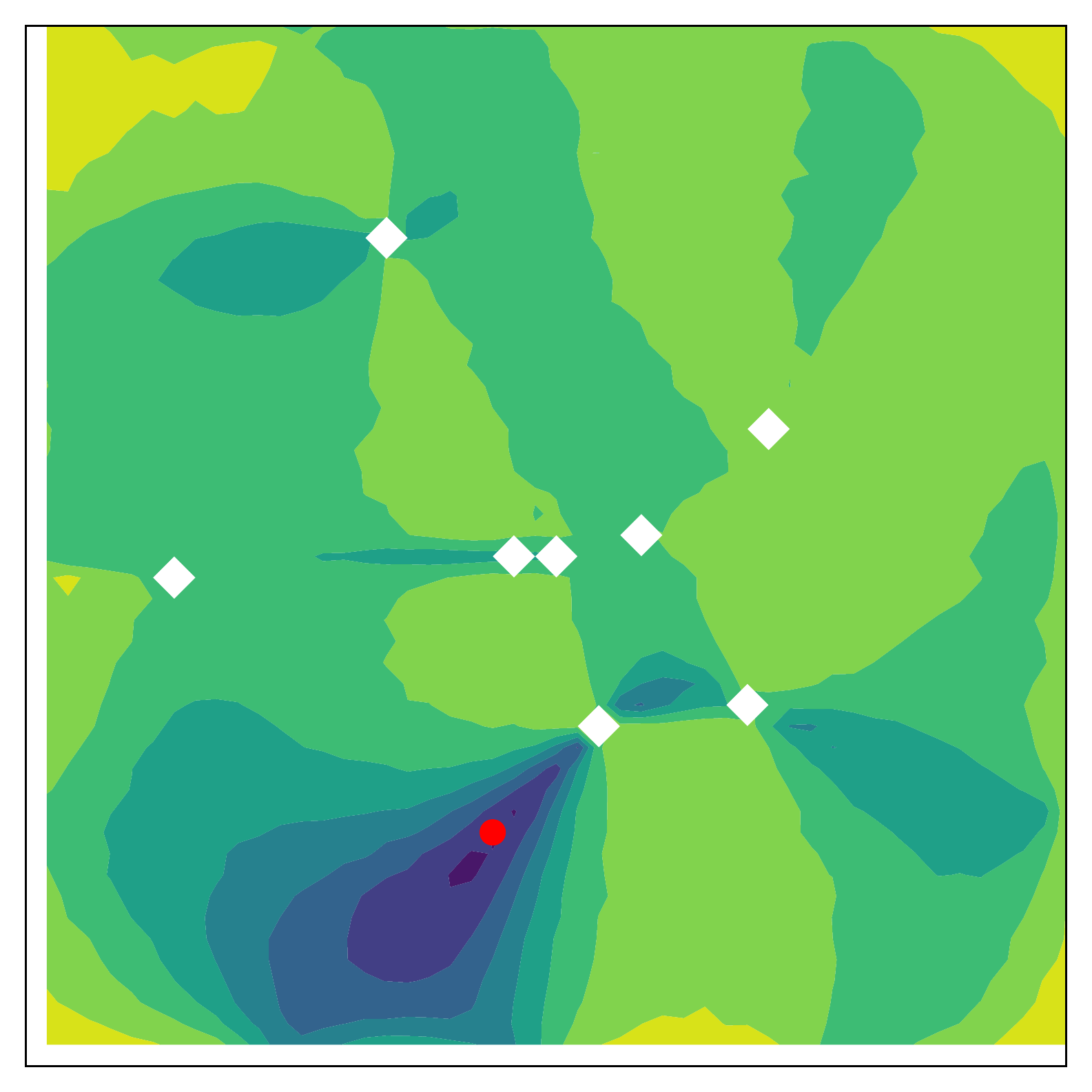}
\caption{Optimal experimental design with the BPN criterion, with $p = \infty$ used.
(The filled contours depict $\text{BPN}(\COexperiment)$ as a function of $t_m$ where the lexicographic ordering of the panels corresponds to $m = 1,\dots,9$.
The white points represent the fixed values of $\{t_i\}_{i=1}^{m-1}$ and the red points represent the minimal value of $t_m$.)
}
\label{fig: pinf}
\end{figure}

For $p=2$ it can be shown that the optimal experiment according to BPN coincides with the optimal experiment according to BDT (see Proposition~\ref{thm: aca equivalence}).
The points $\{t_i\}_{i=1}^m$ were optimised sequentially in Figure~\ref{fig: p2}, for convenience, and we note that a space-filling design was obtained.
For $p = \infty$, in contrast, it is clear from Figure~\ref{fig: pinf} that the optimal design from BPN is no longer space-filling.
However, since it is not easily computed, it is unclear whether or not the optimal experiment according to BPN resembles the optimal experiment according to BDT.

\subsection{Analysis and Open Questions}

Our final aim is to address the question of whether $\COexperimentspace_{\text{BPN}}^{\ast} \stackrel{?}{=} \COexperimentspace_{\text{BDT}}^{\ast}$.
To this end, we first present a positive result in Proposition~\ref{thm: aca equivalence} and then present a negative result in Proposition~\ref{thm: counterexample}.

\subsubsection{A Positive Result}

Under an assumption on the form of the loss $\COloss$, the following equivalence was established in \cite{Cockayne2017}:

\begin{proposition} \label{thm: aca equivalence}
Consider a loss function of the form $\COloss(\COstate , \COstate') = \| \phi(\COstate) - \phi(\COstate') \|_\Phi^2$ where $\phi \colon \COstatespace \rightarrow \Phi$ takes values in an inner product space $\Phi$, with inner product $\langle \cdot , \cdot \rangle_\Phi$ and induced norm $\| \varphi \|_\Phi = \langle \varphi , \varphi \rangle_\Phi^{1/2}$.
Suppose that the first conclusion of Proposition~\ref{prop: mean is Bayes act} holds (i.e.\ Eq.~\eqref{eq: bayes act condition}).
Then $\COexperimentspace_{\text{\emph{BPN}}}^{\ast} = \COexperimentspace_{\text{\emph{BDT}}}^{\ast}$.
\end{proposition}

For completeness, a slightly more concise proof of this result is included in Appendix~\ref{ap: proof}.
This form of loss is commonly encountered and includes the Wasserstein distance with $p=2$ as a particular instance (see Eq.~\eqref{eq: Wass dist}).
A consequence of Proposition~\ref{thm: aca equivalence} is that one can minimise the BPN criterion as an alternative to the BDT criteria in circumstances where the hypotheses of the Proposition hold.

\subsubsection{A Negative Result}

A converse result to Proposition~\ref{thm: aca equivalence} can also be constructed.
The following is inspired by, but slightly more elegant than, a corresponding result in \cite{Cockayne2017}:

\begin{proposition} \label{thm: counterexample}
Suppose that the state space $\COstatespace$ can be partitioned into three disjoint subsets, each with positive probability under $\pi_{\COstaterv}$.
Then there exists a loss function $\COloss$ and a set of candidate experiments $\COexperimentspace$ such that $\COexperimentspace_{\text{\emph{BPN}}}^{\ast} \neq \COexperimentspace_{\text{\emph{BDT}}}^{\ast}$.
\end{proposition}

The proof of this result is constructive and is included in Appendix~\ref{ap: counterexample}.
It is clear that the assumptions on $\COstatespace$ and $\pi_{\COstaterv}$ are weak, to the point of being trivial.

\subsubsection{Open Questions}

An explicit characterisation of the loss functions $\COloss$ for which $\COexperimentspace_{\text{BPN}}^{\ast} = \COexperimentspace_{\text{BDT}}^{\ast}$ is not, at least to our knowledge, available at present.
In particular, the analytic intractability of optimal experiments in all but the simplest of numerical tasks leaves it unclear whether there exist a numerical task of practical importance for which $\COexperimentspace_{\text{BPN}}^{\ast} \neq \COexperimentspace_{\text{BDT}}^{\ast}$.

In general, any utility function $\COutility$ from the Bayesian experimental design literature provides a criterion $\text{BED}(\COexperiment)$ that can be studied from an information-based complexity standpoint.
In particular, the issue of how $\text{BED}(\COexperiment^{\ast})$ scales with $\text{dim}(\COstatespace)$ when $\COexperiment^{\ast} \in \COexperimentspace_{\text{BED}}^{\ast}$, so-called \emph{tractability}, can be studied.

\section{Discussion}

The aim of this article was to build on the earlier work of \cite{Kadane1985}, drawing attention to a wider range of optimality criteria that are used in the Bayesian experimental design literature and considering their use in the numerical context.
In particular, we explained how the competing objectives of parameter estimation and uncertainty quantification lead, in general, to different notions of optimal information for a probabilistic numerical method.

One criterion, called $\text{BPN}$, was explored in detail.
This formalised the idea that probability mass ought to be located close to the true quantity of interest.
However, several other factors are also relevant in the design of a probabilistic numerical method and were not discussed.
Indeed, the $\text{BPN}$ criterion does not encode the notion that a posterior ought to be well calibrated, nor the notion that inferences should be robust to prior mis-specification, nor does it attempt more nuanced control (e.g.\ at the wall clock level) of computational cost \cite{Tuchscherer1983}.
In addition, no attempt was made to extend the notions introduced in this work to the adaptive context, where notions from sequential Bayesian experimental design are needed \cite{Ghosh1991}.
These are active areas of research and we look forward to seeing how these ideas are developed.


\appendix

\section{Proofs}

This appendix collects together proofs for the results quoted in the main text.

\subsection{Proof of Proposition~\ref{prop: norm vs extensive}} \label{ap: norm vs ext proof}

Let $\COdecision_\COexperiment^\circ \in \COdecisionspace_\COexperiment$ be any decision rule such that $\COdecision_\COexperiment^\circ(\COinfo) \in \COactionspace_\COexperiment^{\ast}(\COinfo)$ for all $\COinfo \in \COinfospace$.
Then, for any other $\COdecision_\COexperiment \in \COdecisionspace_\COexperiment$,
\begin{align*}
\text{BR}(\COexperiment,\COdecision_\COexperiment^\circ) - \text{BR}(\COexperiment,\COdecision_\COexperiment) & = \iint \COloss(\COstate , \COdecision_\COexperiment^\circ(\COinfo)) - \COloss(\COstate , \COdecision_\COexperiment(\COinfo) ) \, \mathrm{d}\pi_{\COinforv | \COstate,\COexperiment}(\COinfo) \, \mathrm{d}\pi_{\COstaterv}(\COstate) \\
& = \int \left( \int \COloss(\COstate , \COdecision_\COexperiment^\circ(\COinfo)) - \COloss(\COstate , \COdecision_\COexperiment(\COinfo) ) \, \mathrm{d}\pi_{\COstaterv | \COinfo,\COexperiment}(\COstate) \right) \, \mathrm{d}\pi_{\COinforv | \COexperiment}(\COinfo) \\
& \leq \int 0 \, \mathrm{d}\pi_{\COinforv | \COexperiment}(\COinfo) \; = \; 0.
\end{align*}
Thus $\text{BR}(\COexperiment,\COdecision_\COexperiment^\circ) \leq \text{BR}(\COexperiment,\COdecision_\COexperiment)$ and $\COdecision_\COexperiment^\circ$ is a Bayes rule.
Conversely, let $\COdecision_\COexperiment^{\ast} \in \COdecisionspace_\COexperiment$ be a Bayes rule.
Then
\begin{align*}
0 & \geq \iint \COloss(\COstate,\COdecision_\COexperiment^{\ast}(\COinfo)) - \COloss(\COstate,\COdecision_\COexperiment^\circ(\COinfo) ) \, \mathrm{d}\pi_{\COinforv | \COstate,\COexperiment}(\COinfo) \, \mathrm{d}\pi_{\COstaterv}(\COstate) \\
& = \int \underbrace{\left( \int \COloss(\COstate,\COdecision_\COexperiment^{\ast}(\COinfo)) - \COloss(\COstate,\COdecision_\COexperiment^\circ(\COinfo) ) \, \mathrm{d}\pi_{\COstaterv | \COinfo,\COexperiment}(\COstate) \right)}_{(\ast)} \, \mathrm{d}\pi_{\COinforv | \COexperiment}(\COinfo).
\end{align*}
The definition of $\COdecision_\COexperiment^\circ$ implies that $(\ast) \geq 0$.
Thus $\COloss(\COstate,\COdecision_\COexperiment^{\ast}(\COinfo)) = \COloss(\COstate,\COdecision_\COexperiment^\circ(\COinfo) )$ holds $\pi_{\COinforv | \COexperiment}$-almost everywhere and so $\COdecision_\COexperiment^{\ast}(\COinfo) \in \COactionspace_\COexperiment^{\ast}(\COinfo)$ for $\pi_{\COinforv | \COexperiment}$-almost all $\COinfo \in \COinfospace_\COexperiment$, as required.

\subsection{Proof of Proposition~\ref{prop: mean is Bayes act}} \label{ap: Bayes act proof}

Fix $\COexperiment \in \COexperimentspace$ and $\COinfo_\COexperiment \in \COinfospace_\COexperiment$.
Consider the function
\begin{align*}
f(\COaction) & \COdefeq \int \COloss(\COstate,\COaction) \, \mathrm{d}\pi_{\COstaterv | \COinfo,\COexperiment}(\COstate)
\end{align*}
and recall that the set $\COactionspace_\COexperiment^{\ast}(\COinfo_\COexperiment)$ of Bayes acts is defined as the elements $\COaction \in \COactionspace$ for which $f(\COaction)$ is minimised.
The assumptions on $\COloss$ and $\phi$ imply that $f$ can be twice differentiated.
The first derivative is easily computed:
the integrability assumption on $\phi$ permits differentiation under the integral sign, yielding
\begin{align*}
\frac{\mathrm{d}f}{\mathrm{d}\COaction} & = \int \frac{\mathrm{d}}{\mathrm{d}\COaction} \COloss(\COstate,\COaction) \, \mathrm{d}\pi_{\COstaterv | \COinfo,\COexperiment}(\COstate) \\
& = \int \left\{ - 2 \frac{\mathrm{d}\phi}{\mathrm{d}\COaction}(\COaction)^\top [\phi(\COstate) - \phi(\COaction)] \right\} \, \mathrm{d}\pi_{\COstaterv | \COinfo,\COexperiment}(\COstate) \\
& = - 2 \frac{\mathrm{d}\phi}{\mathrm{d}\COaction}(\COaction)^\top \underbrace{\int [\phi(\COstate) - \phi(\COaction)] \, \mathrm{d}\pi_{\COstaterv | \COinfo,\COexperiment}(\COstate)}_{(\ast)} .
\end{align*}
Since the matrix $\frac{\mathrm{d}\phi}{\mathrm{d}\COaction}$ has full row rank, and since the space $\COactionspace$ is without boundary, if $\COaction$ locally extremises $f$, then the term $(\ast)$ must vanish.
This last condition is precisely Eq.~\eqref{eq: bayes act condition}.

For the converse result, assume that $\pi_{\COstaterv | \COinfo, \COexperiment}$ is a non-atomic distribution on $\COstatespace$, since otherwise the result is trivial.
Let $\frac{\partial^2 \phi}{\partial \COaction_i \partial \COaction_j}$ be the $m$-dimensional column vector with entries $\frac{\partial^2 \phi_k}{\partial \COaction_i \partial \COaction_j}$.
Then $f$ is seen to have continuous second derivative
\begin{align*}
\left[ \frac{\mathrm{d}^2 f}{\mathrm{d} \COaction^2} \right]_{i,j} & = \frac{\partial^2 f}{\partial \COaction_i \partial \COaction_j} \\
& = - 2 \sum_{k=1}^m \frac{\partial^2 \phi_k}{\partial \COaction_i \partial \COaction_j} \int [ \phi_k(\COstate) - \phi_k(\COaction) ] \, \mathrm{d} \pi_{\COstaterv | \COinfo,\COexperiment}(\COstate) + 2 \sum_{k=1}^m \frac{\partial \phi_k}{\partial \COaction_i} \frac{\partial \phi_k}{\partial \COaction_j} \\
& = - 2 \frac{\partial^2 \phi}{\partial \COaction_i \partial \COaction_j}^\top \int [\phi(\COstate) - \phi(\COaction)] \, \mathrm{d}\pi_{\COstaterv | \COinfo,\COexperiment}(\COstate) + 2 \left[ \frac{\mathrm{d}\phi}{\mathrm{d}\COaction}^\top \frac{\mathrm{d}\phi}{\mathrm{d}\COaction} \right]_{i,j}
\end{align*}
since $\phi$ is assumed to be twice continuously differentiable.
The result will follow if $f$ is coercive, since a unique local extremum of a coercive function with continuous second derivatives must be a global minimum of that function.
To this end, consider the reverse triangle inequality
\begin{align*}
f(\COaction) & = \int \|\phi(\COstate) - \phi(\COaction)\|_2^2 \, \mathrm{d}\pi_{\COstaterv | \COinfo,\COexperiment}(\COstate) \\
& \geq \int \left| \|\phi(\COstate)\|_2 - \|\phi(\COaction)\|_2 \right|^2 \, \mathrm{d}\pi_{\COstaterv | \COinfo,\COexperiment}(\COstate) \\
& = \underbrace{\int \| \phi(\COstate) \|_2^2 \, \mathrm{d}\pi_{\COstaterv | \COinfo,\COexperiment}(\COstate)}_{c_1} - 2 \|\phi(\COaction)\|_2 \underbrace{ \int \| \phi(\COstate) \|_2 \, \mathrm{d}\pi_{\COstaterv | \COinfo,\COexperiment}(\COstate) }_{c_2} + \|\phi(\COaction)\|_2^2 \\
& \COqefed g(\|\phi(\COaction)\|_2)
\end{align*}
where $g$ is the quadratic $g(z) = c_1 - 2 c_2 z + z^2$.
From Jensen's inequality, $c_2^2 \leq c_1$ with strict inequality if $\pi_{\COstaterv | \COinfo, \COexperiment}$ is not an atomic distribution on $\COstatespace$.
Thus, the coercivity of $\phi$ implies the coercivity of $f$, and the proof is complete.

\subsection{Proof of Proposition~\ref{thm: aca equivalence}} \label{ap: proof}

To start, we indicate how the assumptions on the loss function will be used.
Observe that
\begin{equation}
\label{eq:use_inner_prod}
\iint \| \phi(\COstate) - \phi(\COstate') \|_\Phi^2 \, \mathrm{d} \pi_{\COstaterv | \COinfo , \COexperiment}(\COstate) \, \mathrm{d} \pi_{\COstaterv | \COinfo , \COexperiment}(\COstate') = 2 \int \| \phi(\COstate) - \bar{\phi}_{\COinfo, \COexperiment} \|_\Phi^2 \, \mathrm{d} \pi_{\COstaterv | \COinfo , \COexperiment}(\COstate)
\end{equation}
where $\bar{\phi}_{\COinfo , \COexperiment} = \int \phi(\COstate) \, \mathrm{d} \pi_{\COstaterv | \COinfo , \COexperiment}(\COstate)$.
Indeed,
\[
\phi(\COstate) - \phi(\COstate') = (\phi(\COstate) - \bar{\phi}_{\COinfo , \COexperiment}) - (\phi(\COstate') - \bar{\phi}_{\COinfo , \COexperiment})
\]
and
\[
\|\phi(\COstate) - \phi(\COstate')\|_\Phi^2 = \|\phi(\COstate) - \bar{\phi}_{\COinfo , \COexperiment}\|_\Phi^2 - 2 \langle \phi(\COstate) - \bar{\phi}_{\COinfo , \COexperiment} , \phi(\COstate') - \bar{\phi}_{\COinfo , \COexperiment} \rangle_\Phi + \|\phi(\COstate') - \bar{\phi}_{\COinfo , \COexperiment}\|_\Phi^2 ,
\]
where the first and third terms have identical integrals under $\mathrm{d} \pi_{\COstaterv | \COinfo , \COexperiment}(\COstate) \, \mathrm{d} \pi_{\COstaterv | \COinfo , \COexperiment}(\COstate')$ and
\[
\int \langle \phi(\COstate) - \bar{\phi}_{\COinfo , \COexperiment} , \phi(\COstate') - \bar{\phi}_{\COinfo , \COexperiment} \rangle_\Phi \, \mathrm{d} \pi_{\COstaterv | \COinfo , \COexperiment}(\COstate) = \langle 0 , \phi(\COstate') - \bar{\phi}_{\COinfo , \COexperiment} \rangle_\Phi = 0.
\]

Next, we use the fact that $\mathrm{d}\pi_{\COinforv | \COstate, \COexperiment}(\COinfo) \, \mathrm{d}\pi_{\COstaterv}(\COstate) = \mathrm{d}\pi_{\COstaterv | \COinfo , \COexperiment}(\COstate) \, \mathrm{d}\pi_{\COinforv | \COexperiment}(\COinfo)$ to observe that, for the loss $\COloss(\COstate , \COstate') = \|\phi(\COstate) - \phi(\COstate')\|_\Phi^2$,
\begin{align}
& \text{BPN}(\COexperiment) \notag \\
& \quad = \iiint \|\phi(\COstate) - \phi(\COstate')\|_\Phi^2 \, \mathrm{d} \pi_{\COstaterv | \COinfo, \COexperiment}(\COstate') \, \mathrm{d} \pi_{\COinforv | \COstate, \COexperiment}(\COinfo) \, \mathrm{d} \pi_{\COstaterv}(\COstate) \qquad \text{(from Eq.~\eqref{eq: proposed})} \notag \\
& \quad = \int \left( \iint \| \phi(\COstate) - \phi(\COstate') \|_\Phi^2 \, \mathrm{d} \pi_{\COstaterv | \COinfo , \COexperiment}(\COstate) \, \mathrm{d} \pi_{\COstaterv | \COinfo , \COexperiment}(\COstate') \right) \, \mathrm{d} \pi_{\COinforv | \COexperiment}(\COinfo) \label{eq: fubini} \\
& \quad = \int \left( 2 \int \| \phi(\COstate) - \bar{\phi}_{\COinfo, \COexperiment} \|_\Phi^2 \, \mathrm{d} \pi_{\COstaterv | \COinfo , \COexperiment}(\COstate) \right) \, \mathrm{d} \pi_{\COinforv | \COexperiment}(\COinfo) \qquad \text{(from Eq.~\eqref{eq:use_inner_prod})} . \label{eq: C exp final}
\end{align}

Third, we recall that, for the loss $\COloss(\COstate , \COstate') = \|\phi(\COstate) - \phi(\COstate')\|_\Phi^2$, a Bayes act $\COaction \in \COactionspace$ must satisfy $\phi(\COaction) = \bar{\phi}_{\COinfo , \COexperiment}$ under the first conclusion of Proposition~\ref{prop: mean is Bayes act} (i.e.\ Eq.~\eqref{eq: bayes act condition}).
Therefore,
\begin{align}
\text{BR}(\COexperiment,\COdecision_\COexperiment^{\ast}) & = \iint \| \phi(\COstate) - \bar{\phi}_{\COinfo , \COexperiment} \|_\Phi^2 \, \mathrm{d}\pi_{\COinforv | \COstate , \COexperiment}(\COinfo) \, \mathrm{d}\pi_{\COstaterv}(\COstate) \qquad \text{(from Eq.~\eqref{eq: risk})} \nonumber \\
& = \iint \| \phi(\COstate) - \bar{\phi}_{\COinfo , \COexperiment} \|_\Phi^2 \, \mathrm{d}\pi_{\COstaterv | \COinfo , \COexperiment}(\COstate) \, \mathrm{d}\pi_{\COinforv | \COexperiment}(\COinfo) , \label{eq: risk exp final}
\end{align}
where the final equality follows the same argument used to obtain Eq.~\eqref{eq: fubini}.
This final expression, in Eq.~\eqref{eq: risk exp final}, is observed to be exactly half of the expression for $\text{BPN}(\COexperiment)$ obtained in Eq.~\eqref{eq: C exp final}.
Therefore, minimisation of Eq.~\eqref{eq: risk exp final} is equivalent to minimisation of Eq.~\eqref{eq: C exp final} and it follows that $\COexperimentspace_{\text{BDT}}^{\ast} = \COexperimentspace_{\text{BPN}}^{\ast}$, as claimed.

\subsection{Proof of Proposition~\ref{thm: counterexample}} \label{ap: counterexample}

To start, suppose we are presented with a state space $\COstatespace$, that can be partitioned into three measurable subsets $\COsoneintext$, $\COstwointext$, and $\COsthreeintext$, with respective strictly positive probabilities $\pi_\COsone$, $\pi_\COstwo$, and $\pi_\COsthree$.
Without loss of generality we assume that $0 < \pi_\COsone \leq \pi_\COstwo \leq \pi_\COsthree < 1$, noting that $\pi_\COsone + \pi_\COstwo + \pi_\COsthree = 1$.
In what follows we exhibit a collection of candidate experiments $\COexperimentspace$ that, in effect, reveal some information about the state $\COstate$ at the coarse level of the partition $\{\COsoneintext,\COstwointext,\COsthreeintext\}$.
Let $\COsuit \colon \COstatespace \rightarrow \{\COsoneintext,\COstwointext,\COsthreeintext\}$ be the map that assigns each $\COstate \in \COstatespace$ to its partition element.
Consider a set of just two candidate experiments, in each case reporting a deterministic observation of the latent state:
\begin{align*}
\COexperimentspace & = \left\{ \begin{array}{ll} \COexperiment_1 \; : & \COinfo(\COstate) = 1_{\COsuit(\COstate) \in \{\COsone\}} \\ \COexperiment_2 \; : & \COinfo(\COstate) = 1_{\COsuit(\COstate) \in \{\COsone , \COstwo\}} \end{array} \right\}.
\end{align*}
The aim is to decide whether experiment $\COexperiment_1$ or $\COexperiment_2$ should be performed.
The informativeness of an experiment will be measured as described in the main text, based on the so-called 0-1 loss
\begin{align*}
\COloss(\COstate,\COstate') & = \begin{cases} 0 & \text{if } 1_{\COsuit(\COstate) = \COsone} = 1_{\COsuit(\COstate') = \COsone} , \\ 1 & \text{otherwise.} \end{cases}
\end{align*}
The 0-1 loss depends on the state $\COstate$ only through the indicator function $1_{\COsuit(\COstate) = \COsone}$, so that our task is equivalent to guessing whether the true $\COsuit(\COstate)$ is $\COsoneintext$ or not, based on information obtained in the experiment.
Thus, in a small abuse of notation, we may re-define the action space to be $\COactionspace = \{\COsoneintext, \neg \COsoneintext\}$.

\vspace{5pt}
\noindent \textbf{Bayesian Decision Theory}
First we derive the BDT-optimal experiment(s) $\COexperimentspace_{\text{BDT}}^{\ast}$.
Without loss of generality we can restrict attention, in the search for a Bayes rule, to the set of non-random decision rules of the form $\COdecision_{\COexperiment} (\COinfo) = 1_{\COinfo = 0} \COaction_0 + 1_{\COinfo = 1} \COaction_1$ for some $\COaction_1, \COaction_2 \in \COactionspace$ to be specified.
The actions $\COaction_0, \COaction_1$ are selected to minimise the risk
\begin{align*}
\text{BR}(\COexperiment,\COdecision_\COexperiment) & = \int \bigl[ \COinfo(\COstate) \COloss(\COstate, \COaction_1) + (1 - \COinfo(\COstate)) \COloss(\COstate, \COaction_0) \bigr] \, \mathrm{d}\pi_{\COstaterv}(\COstate) .
\end{align*}
Inputting the form of each experiment, we obtain
\begin{align*}
\text{BR}(\COexperiment_1,\COdecision_{\COexperiment_1}) & = \pi_\COsone 1_{\COaction_1 \neq \COsone} + \pi_\COstwo 1_{\COaction_0 = \COsone} + \pi_\COsthree 1_{\COaction_0 = \COsone} \\
\text{BR}(\COexperiment_2,\COdecision_{\COexperiment_2}) & = \pi_\COsone 1_{\COaction_1 \neq \COsone} + \pi_\COstwo 1_{\COaction_1 = \COsone} + \pi_\COsthree 1_{\COaction_0 = \COsone} .
\end{align*}
The expression for $\text{BR}(\COexperiment_1,\COdecision_{\COexperiment_1})$ is minimised by $\COaction_1 = \neg \COsoneintext$ and $\COaction_0 = \neg \COsoneintext$, since $\pi_\COsone, \pi_\COstwo , \pi_\COsthree > 0$.
Similarly, the expression for $\text{BR}(\COexperiment_2,\COdecision_{\COexperiment_2})$ is minimised by $\COaction_1 = \neg \COsoneintext$ and $\COaction_0 = \neg \COsoneintext$, since $\pi_\COsone \leq \pi_\COstwo$ and $\pi_\COsthree > 0$.
It follows that $\text{BR}(\COexperiment_2,\COdecision_{\COexperiment_2}^{\ast}) = \text{BR}(\COexperiment_1,\COdecision_{\COexperiment_1}^{\ast}) = \pi_\COsone$ and both experiments are optimal.
Therefore, $\COexperimentspace_{\text{BDT}}^{\ast} = \{\COexperiment_1 , \COexperiment_2\}$.

\vspace{5pt}
\noindent \textbf{Bayesian Experimental Design}
Now we derive the BPN-optimal experiment(s) $\COexperimentspace_{\text{BPN}}^{\ast}$.
Through direct computation we see that
\begin{align*}
\text{BPN}(\COexperiment_1) & = \pi_\COsone \int 1_{\COsuit(\COstate') \neq \COsone} \, \mathrm{d}\pi_{\COstaterv | \COinfo = 1, \COexperiment_1}(\COstate') + \pi_\COstwo \int 1_{\COsuit(\COstate') = \COsone} \, \mathrm{d}\pi_{\COstaterv | \COinfo = 0, \COexperiment_1}(\COstate') \nonumber \\
& \phantom{=} \quad  + \pi_\COsthree \int 1_{\COsuit(\COstate') = \COsone} \, \mathrm{d}\pi_{\COstaterv | \COinfo = 0, \COexperiment_1}(\COstate') \\
& = \pi_\COsone \times 0 + \pi_\COstwo \times 0 + \pi_\COsthree \times 0 \; = \; 0 \\
\text{BPN}(\COexperiment_2) & = \pi_\COsone \int 1_{\COsuit(\COstate') \neq \COsone} \, \mathrm{d}\pi_{\COstaterv | \COinfo = 1, \COexperiment_2}(\COstate') + \pi_\COstwo \int 1_{\COsuit(\COstate') = \COsone} \, \mathrm{d}\pi_{\COstaterv | \COinfo = 1, \COexperiment_2}(\COstate') \nonumber \\
& \phantom{=} \quad + \pi_\COsthree \int 1_{\COsuit(\COstate') = \COsone} \, \mathrm{d}\pi_{\COstaterv | \COinfo = 0, \COexperiment_2}(\COstate') \\
& = \pi_\COsone \frac{\pi_\COstwo}{\pi_\COsone + \pi_\COstwo} + \pi_\COstwo \frac{\pi_\COsone}{\pi_\COsone + \pi_\COstwo} + \pi_\COsthree \times 0 \; = \; \frac{2 \pi_\COsone \pi_\COstwo}{\pi_\COsone + \pi_\COstwo} .
\end{align*}
Since $\pi_\COstwo \geq \pi_\COsone > 0$, it follows that experiment $\COexperiment_1$ is the \emph{only} optimal experiment.
Therefore, $\COexperimentspace_{\text{BPN}}^{\ast} = \{\COexperiment_1\} \neq \{\COexperiment_1 , \COexperiment_2\} = \COexperimentspace_{\text{BDT}}^{\ast}$, as claimed.


\paragraph{Acknowledgements:}
The authors are grateful to \textcolor{black}{an anonymous reviewer for their thoughtful comments}, to Dave Woods for discussion of this work, as well as to the organisers and participants of the RICAM workshop on \textit{Multivariate Algorithms and Information-Based Complexity}, for which this paper was prepared.

CJO and MG were supported by the Lloyd's Register Foundation programme on Data-Centric engineering at the Alan Turing Institute, UK.
TJS was supported by the Excellence Initiative of the German Research Foundation (DFG) through the Freie Universit\"at Berlin.
MG was supported by the EPSRC grants [EEP/P020720/1, EP/R018413/1, EP/R034710/1, EP/R004889/1] and a Royal Academy of Engineering Research Chair.


\end{document}